\documentclass[pdftex,twocolumn,epjc3]{svjour3}      

\RequirePackage[T1]{fontenc}

\smartqed  

\RequirePackage[american]{babel} 
\RequirePackage{booktabs} 
\RequirePackage{textcomp} 
\RequirePackage{graphicx}
\RequirePackage{siunitx}
\RequirePackage{flushend}
\RequirePackage[numbers,sort&compress]{natbib}
\RequirePackage[colorlinks,citecolor=blue,urlcolor=blue,linkcolor=blue]{hyperref}
\RequirePackage{upgreek}
\RequirePackage{amsmath}
\RequirePackage{amssymb}
\RequirePackage{bm}

\journalname{Eur. Phys. J. E}

\newcommand{\df}{\mathrm{d}}
\newcommand{\vv}[1]{\bm{#1}}
\newcommand{\te}[1]{\mathrm{#1}}

\newcommand{\mups}{\mu_{\rm ps}}
\newcommand{\muau}{\mu_{\rm au}}
\newcommand{\vslip}{v_{\te s}}
\newcommand{\phipa}{\phi_{\te{pa}}}
\newcommand{\phiap}{\phi_{\te{ap}}}

\newcommand{\av}[1]{\left\langle #1 \right\rangle}

\newcommand{\Tint}{\widetilde{T}}

\newcommand{\ie}{i.e.}

\newcommand{\Dr}{{D_\te r}}

\begin{document}

\title{Thermotaxis of Janus Particles\thanksref{t1}}


\author{
  Sven Auschra\thanksref{e1,addr1}
  \and
  Andreas Bregulla\thanksref{e2,addr2}
  \and
  Klaus Kroy\thanksref{e3,addr1}
  \and
   Frank Cichos\thanksref{e2,addr2}         
}

\thankstext[$\star$]{t1}{%
  Contribution to the Topical Issue “Motile Active Matter”, edited by Gerhard Gompper, Clemens Bechinger, Roland G. Winkler, Holger Stark
}
\thankstext{e1}{e-mail: sven.auschra@itp.uni-leipzig.de}
\thankstext{e2}{e-mail: cichos@uni-leipzig.de}
\thankstext{e3}{e-mail: klaus.kroy@uni-leipzig.de}

\institute{Institute for Theoretical Physics, Leipzig University, 04103 Leipzig, Germany\label{addr1} 
\and
Peter Debye Institute for Soft Matter Physics, Leipzig University, 04103 Leipzig, Germany\label{addr2}
}

\date{Received: date / Accepted: date}

\maketitle

\begin{abstract}
%

The interactions of autonomous microswimmers play an important role for the formation of collective states of motile active matter.  We study them in detail for the common microswimmer-design of two-faced Janus spheres with hemispheres made from different materials. Their chemical and physical surface properties may be tailored to fine-tune their mutual attractive, repulsive or aligning behavior.
To investigate these effects systematically, we monitor the dynamics of a single gold-capped Janus particle in the external temperature field created by an optically heated metal nanoparticle. We quantify the orientation-dependent repulsion and alignment of the Janus particle and explain it in terms of a simple theoretical model for the induced thermoosmotic surface fluxes. The model reveals that the particle's angular velocity is solely determined by the temperature profile on the equator between the  Janus particle's hemispehres and their  phoretic mobility contrast. The distortion of the external temperature field by their heterogeneous heat conductivity is moreover shown to break the apparent symmetry of the problem. 
\end{abstract}

\section{Introduction}

Ranging from flocks of birds via schools of fish to colonies of insects, a distinctive trait displayed by the individual constituents of motile active matter \cite{ramaswamy2010StatMechActMat,menon2010ActMatt,cates2012MicroiolStatMech} is a unique capability to adapt to  environmental cues \cite{vicsek1995SelfDrivenParticles}. Down to the microbial level where all kinds of ``animalcules'' \cite{gest2004discoveryOfMicroorganisms} struggle to locomote through liquid solvents \cite{Purcell77,Lauga2011LifeTheorem}, interactions with boundaries and neighbors and the sensing of chemical gradients \cite{adler1966chemotaxis} are key features involved in the search of food, suitable habitats or mating partners.
Inspired by nature, scientist designed synthetic, inanimate microswimmers that mimic the characteristics of biological swimmers and are more amenable to a systematic investigation of their interactions. A very popular design exploits self-phoresis \cite{Anderson1989} for which numerous experimental and theoretical studies are available \cite{zoettl2016EmergBehav,romanczuk2012ABP,poon2013Review}. Such self-phoretic propulsion relies on the interfacial stresses arising at the particle--fluid interface in the self-generated gradient of an appropriate field (temperature, solute concentration, electrostatic potential). On a coarse-grained hydrodynamic level, this effect is captured by an effective tangential slip of the fluid along the particle surface \cite{Anderson1989} that drives the self-propulsion of the swimmer. Accordingly, thermophoresis \cite{Wurger2007ThermophoresisForces,Fayolle2008ThermophoresisParticles,wuerger2010Review}, diffusiophoresis \cite{golestanian2005Diffusophoresis,Julicher2009Colltrans,Wurger2015Self-DiffusiophoresisMixtures} and electrophoresis \cite{squires2004Electroosmosis,squires2006Electrophoresis} can deliberately be exploited for (or may inadvertently contribute to) the swimming of Janus particles \cite{Jiang2010,Bregulla2014,selmke2018PhotonNudgingI,selmke2018PhotonNudgingII,popescu2016SelfDiffPhor,golestanian2007DesigningSwimmers,bazant2010electrokonPhen,gangwal2008Electrophoresis,moran2010electrophoresis}. The classical Janus-particle design consists of a spherical colloid with hemispheres of distinct physico-chemical properties, which define a polar symmetry axis. Due to the broken symmetry, one expects the axis to align with an external field gradient \cite{Bickel2014Polarization}, the direction being determined by the precise surface properties and the chosen solvent  \cite{saha2019PairingWaltzingScattering,pohl2014ChemotactCollapse}.  The reorientation of microswimmers in external fields is often referred to as \emph{taxis} and has been studied for various phoretic propulsion mechanisms \cite{boymelgreen2012electrophor,Lozano2016Phototaxis,Gomez2017Directionality,Jin2017,Geiseler2017Self-PolarizingWaves,uspal2019LightActivatedParticles}.

What we call an external field can be understood as a template for the influence of container walls or neighboring microswimmers \cite{liebchen2019DomInteract,popescu2019CommentWhichInteract,zoettl2016EmergBehav,popescu2018Chemotaxis} that are at the core of the rich collective phenomena emerging in active fluids \cite{golestanian2012Collbehav,saha2014ClustersAtsersOscill,Husain2017emergStruct,Liebchen2017DynClust,Liebchen2015Cluster,Theurkauff2012Clustering,Kaiser2012Capture,Pohl2015Collapse,Cates2015MIPS}.
That microswimmers are constantly exchanging linear and angular momentum with the ambient fluid generally renders their apparent mutual interactions non-reciprocal.  Next to the thermodynamic field gradients also the hydrodynamic flow field generated by one swimmer at the position of another one affects the swimmers' interactions \cite{elgeti2015Review,zoettl2016EmergBehav}. Generally, interactions mediated by hydrodynamic flow fields \cite{takagi2014HydrOrbits,varma2018Clustering,dunkel2010Scattering,alexander2008Scattering,martinez2018HydBoundStates,berke2008HydrAttr}, optical shadowing \cite{uspal2019LightActivatedParticles,Cohen2014OpticallyDriven} and chemical or optical patterns \cite{Volpe2011MicroswimmersEnvironments,palagi2016lightstruct,Simmchen2016Topograph} may have to be considered, and which of these contributions dominate the observed motion of microswimmers has recently been under debate \cite{liebchen2019DomInteract,popescu2019CommentWhichInteract,liebchen2019ResponseToComment}.

In the present contribution, we report results from an experiment designed to allow for direct measurements of the induced polarization and motion of a single passive (i.e., not self-driven) Janus particle in the temperature field emanating from a localized heat source in its vicinity. In other words, it enables us to single out the passive phoretic response of a self-thermophoretic swimmer to an external temperature field, without having to bother with the autonomous motion that would result from a direct (laser-)heating of the swimmer itself. Typically such measurements are difficult to conduct since pairwise collisions are rare at low concentrations and hard to discern among interfering the many-body effects, at high concentrations. However, by adopting the technique of photon nudging \cite{Qian2013,Bregulla2014} we can direct an individual Janus particle into the vicinity of the local heat source \cite{selmke2018PhotonNudgingI,selmke2018PhotonNudgingII}, without imposing potentially perturbing external fields. This allows us to precisely record the polarization effects and thereby characterize the elusive phoretic repelling and aligning interactions with good accuracy and good statistics. 
Our experimental results are substantiated by a theoretical model that addresses the thermophoretic origin of the interactions and complements recent calculations of the phoretic interactions between two chemically active particles  \cite{nasouri2020PhoreticInteractions2Particles} and the axis-symmetric interactions between two diffusiophoretic Janus particles \cite{nasouri2020AxissymmInteract2JPs}.

\section{Experimental Setup}
\label{sec:experimental}
\begin{figure} \includegraphics[width=1\columnwidth]{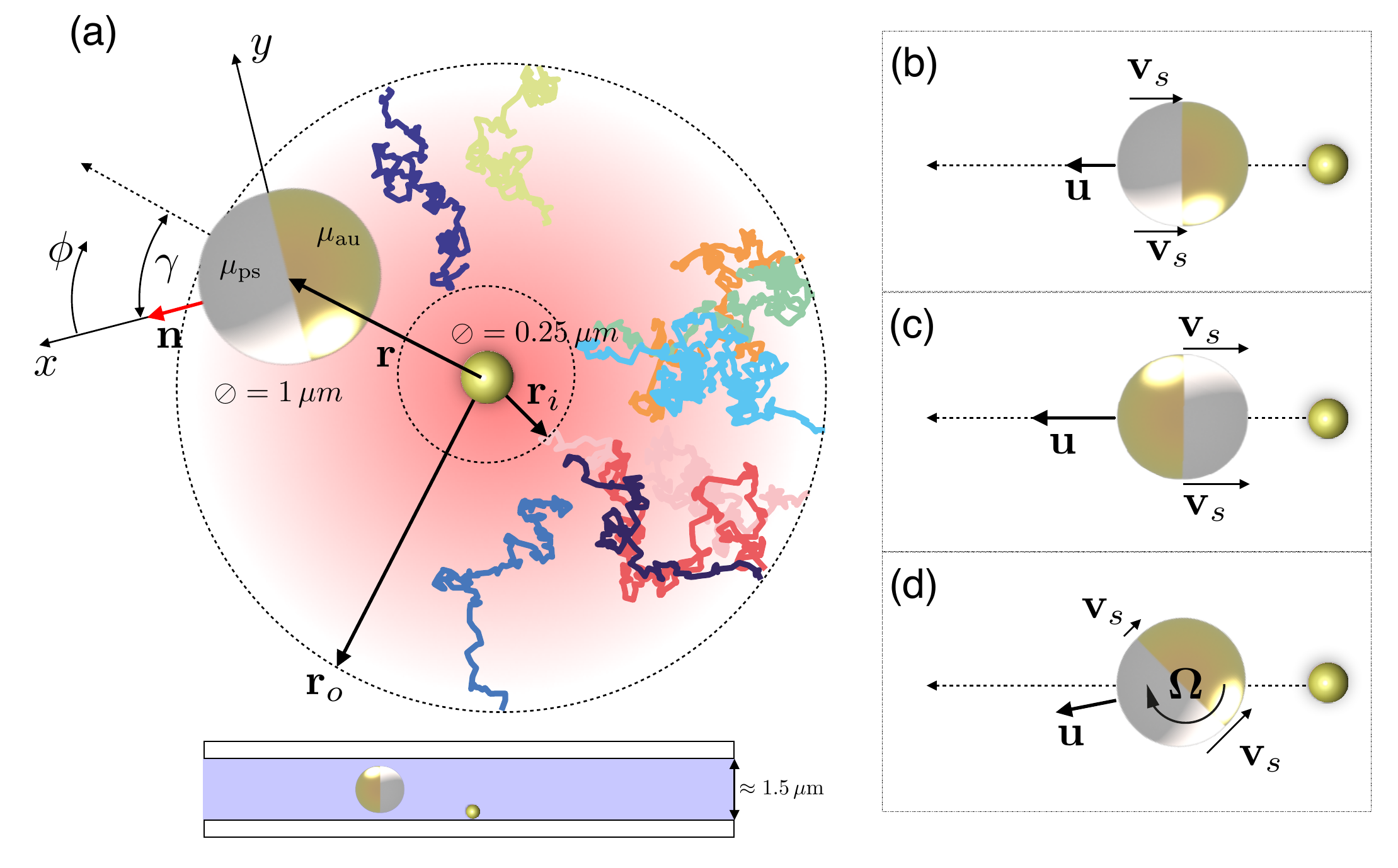}
  \caption{%
  \textbf{Schematic of the experimental setup and the phoretic motion.}
    \textbf{(a)} A passive Janus polystyrene (ps) bead with a thin gold (au) cap (thermophoretic mobilities  $\mups$, $\muau$) is exposed to the temperature gradient around a laser-heated immobilized gold nanoparticle. Particle translation is restricted to the sample plane due to the thin liquid film thickness ($\SI{1.5}{\micro\metre}$). The coordinate frame attached to the particle's geometric center has its $x$-axis aligned with the particle's symmetry axis and pointing towards its ps-side, while  the $z$-axis points into the paper plane. 
The in-plane angle $\phi$ and normal  angle $\theta$ are measured with respect to the $x$-axis and $z$-axis, respectively, and technically (yet not with respect to the particle's polarity) take the role of what is conventionally called ``azimuthal'' and ``polar'' angles, respectively. The orientation angle of the swimmer relative to the heat source is $\gamma$. \textbf{(b--d)} Phoretic translational and rotational velocities $\vv u$, $\vv \Omega$, arise from slip fluxes with velocities $\vv v_\te{s}$ induced by the temperature gradient, chiefly near the particle equator. Arrow lengths and orientations indicate the magnitude and direction of the velocities.}
  \label{fig:method}
\end{figure}

We experimentally explore the interaction of a $\SI{1}{\micro\metre}$ diameter Janus particle with a $\SI{50}{\nano\metre}$ thin gold cap with the temperature field generated by an immobilized $\SI{250}{\nano\metre}$ gold nano-particle optically heated  by a focused laser (wavelength \SI{532}{\nano\metre}). To confine the Janus particle to the vicinity of the heat source, we employ  the feedback control technique of photon nudging \cite{Qian2013,Bregulla2014} that exploits its autonomous motion to steer it to a chosen target. As illustrated in Fig. \ref{fig:method} (a), the steering is only activated when the Janus particle leaves an outer radius $r_{\rm o}$ around the heat source until it has migrated back across an inner radius $r_{\rm i}$, followed by a waiting time of $10$ rotational diffusion times $\tau_{\te r}$  to allow for the decay of orientational biases and correlations \cite{Auschra2020ActivityFieldsPRE,Soeker2020ActivityFieldsPRL}.

 All data recording and feedback is carried out in a custom-made dark field microscopy setup with an inverse frame rate and exposure time of $\SI{5}{\milli\second}$.
Further details regarding the sample preparation, the experimental setup, and the position and orientation analysis are contained in \ref{sec:prep-janus-part}--\ref{sec:particle_tracking}. 
The temperature increment $\Delta T = 
\SI{12}{\kelvin}$  of the heated gold nanoparticle relative to the ambient temperature ($T_0 = \SI{295}{\kelvin}$) is known from a separate measurement using the nematic/isotropic phase transition of a liquid crystal (see \ref{sec:meas-temp-prof}). We  account for the direct influence of the heating laser on the Janus particle and the phoretic velocities, as detailed in \ref{sec:infl-laser-heat}.

\section{Results and Discussion}

\subsection{Theory}
\label{sec:theory}

On the hydrodynamic level of description, the temperature gradient
\(
\vv \nabla_{\parallel}T
\equiv
(\vv I - \vv e_r \vv e_r) \vv \nabla T
\)
along the surface of the Janus particle induces a proportionate interfacial creep flow \cite{derjaguin1987SurfaceForeces,bregulla2016Thermoosmosis}, where $\vv e_r$ denotes the unit vector normal to the particle surface and $\vv I$ the unit matrix. Since the interfacial flow is localized near the particle surface, it is  conveniently represented as a slip boundary condition with slip velocity \cite{Anderson1989,golestanian2007DesigningSwimmers,Bickel2014Polarization} 
\begin{equation}\label{eq:slip_velocity}
  \vv v_{\te{s}}\left(\theta,\phi \right)
  =
  \mu\left(\theta,\phi\right)
  \vv \nabla _{\parallel}T\left(\theta,\phi\right).
\end{equation}
The particle surface is parametrized in terms of the in-plane and normal angles $\phi$ and $\theta$,  as sketched in Fig.~\ref{fig:method} (a,b).  They technically take the role of "azimuthal" and "polar" angles,  respectively, although these notions are not associated with the particle's polar symmetry, here. And $\mu\left(\theta,\phi\right)$ is a phoretic mobility characterizing the varying strength of the creep flow due to the distinct interfacial interactions with the solvent \cite{Bickel2014Polarization}.
The resulting translational propulsion velocity $\vv u$ and the angular velocity $\bm \Omega$
of the Janus particle of radius $a$ are given by  averages over its surface $\mathcal S$: \cite{anderson1991DiffPhoresis,Bickel2014Polarization}
\begin{align}
  \label{eq:particle_velocity}
  \vv u
  &=
    - \frac{1}{4 \pi a^2}
    \oint_{\mathcal S} \df S \, \vv v_{\rm{s}},
  \\[0.5em]
  \label{eq:particle_angularvelocity}
  \vv \Omega
  &=
    -\frac{1}{4 \pi a^2}
    \oint_{\mathcal S} \df \vv S
    \times
    \frac{3}{2a} \vv v_{\rm{s}}.
\end{align}

Further analysis of Eqs.~\eqref{eq:particle_velocity} and \eqref{eq:particle_angularvelocity} becomes possible by the experimental observation that the Janus particle is preferentially aligned with the sample plane. This effect is presumably mostly due to the hydrodynamic flows induced by the heterogeneous heating in the narrow fluid layers between the particle and the glass cover slides \cite{uspal2015ParticlesNearWall,Simmchen2016TopographicalMicroswimmers,Das2015BoundariesSpheres}. 
For simplicity, the following analysis assumes perfect in-plane alignment, thereby neglecting weak perturbations due to  rotational Brownian motion and the weak bottom-heaviness of the Janus particle   \cite{rashidi2020influCapWeight}.
For any given temperature profile $T(\phi,\theta)$ at the surface of the Janus sphere, the components of the translational and rotational velocity can then be expressed as 
    \begin{align}
      \nonumber
      u_x
      &=
        -
        \frac{1}{\pi a}        
        \int_0^{2\pi} \df \phi \,
        \mu(\phi)
        \cos\phi        
        \av{T \sin \theta}_{\theta}(\phi)        
      \\
      \label{eq:theory_ux}
      &\hspace{0.4cm}+
        \frac{\mups - \muau}{2 \pi a}
        \left[ 
        \av{ \frac{T}{\sin\theta} }_{\theta}
        \biggr\rvert_{\phi=\frac{\pi}{2}}
        +
        \av{ \frac{T}{\sin\theta} }_{\theta}
        \biggr\rvert_{\phi=\frac{3\pi}{2}}
        \right],
      \\[0.5em]
      \label{eq:theory_uy}
      u_y
      &=
        -\frac{1}{\pi a}
        \int_0^{2\pi} \df \phi \,
        \mu(\phi)
        \sin\phi        
        \av{T \sin \theta}_{\theta}(\phi),
      \\[0.5em]
      \label{eq:theory_omega}
      \Omega_z
      &=
        \frac{3}{4 \pi a^2}        
        (\mups - \muau)
        \left[
        \av{T}_{\theta}|_{\phi=\frac{3\pi}{2}} - \av{T}_{\theta}|_{\phi=\frac{\pi}{2}}
        \right],
    \end{align}
    where we have introduced the average over the normal angle
    \(
    \av{\bullet}_{\theta}
    \equiv
    \frac12 \int_0^\pi \df \theta ~
    \sin\theta (\bullet)
    \).
    All other velocity components give zero contributions, as the detailed derivation of Eqs.~\eqref{eq:theory_ux}--\eqref{eq:theory_omega} in \ref{sec:derivation_theory} shows.

Motivated by theoretical studies of chemotactic active colloids \cite{saha2019PairingWaltzingScattering}, we further employ the following model for the angular velocity:     
    \begin{equation}
      \label{eq:Omega_goles}
      \Omega_z(\gamma)
      =
      \Omega_1 \sin\gamma
      +
      \Omega_2 \sin(2\gamma),
    \end{equation}
    introducing the independent parameters $\Omega_{1,2}$. This is a natural extension of $\Omega_z \propto (\mups - \muau)\sin\gamma$ for particles with isotropic heat conductivity \cite{Bickel2014Polarization}, to account for the material heterogeneities of the Janus sphere. 
    Similar (reflection) methods as presented in \cite{saha2019PairingWaltzingScattering, boymelgreen2012electrophor,boymelgreen2012acDielecJP} might be employed to establish a connection between  $\Omega_{1,2}$ and the material and interaction parameters of the Janus particle and the ambient fluid, but we do not pursue this further, here. The crucial feature is that the term $\propto \sin(2\gamma)$ acknowledges the higher periodicity of the effect of the two hemipsheres' distinct heat conductivities onto the rotational motion. 

The competition between the phoretic alignment of the Janus particle and its orientational dispersion by rotational diffusion can be described by the Fokker--Planck equation \cite{doi86,golestanian2012Collbehav}
\begin{equation}
    \label{eq:FPE_full}
    \partial_t f
    =
    -
    \vv{\mathcal R}
    \cdot
    \left(
        \vv \Omega - \Dr \vv{\mathcal R}
    \right) f,
  \end{equation}
  for the dynamic probability density $f(t,\vv n)$ to find the particle at time $t$ with an orientation $\vv n$ (relative to the heat source).
  The rotational operator
  \(
  \vv{\mathcal R} 
  \equiv 
  \vv n \times \vv \nabla_n
  \)  
  includes the nabla operator $\vv \nabla_n$ with respect to the particle's orientational degrees of freedom, and $D_\te{r}$ denotes the (effective \cite{rings2012rotHBM,falasco2016HBM}) rotational diffusion coefficient.  With the mentioned approximation of a strict in-plane orientation of the particle axis, Eq.~\eqref{eq:FPE_full} greatly simplifies \cite{Bickel2014Polarization} to
  \(
  \partial_t f = -\partial_\gamma J,
  \)
with the flux
\begin{equation}
  \label{eq:J}
  J(\gamma,t)
  =
  - \Omega_z(\gamma) f(\gamma,t) 
  - 
  \Dr \partial_\gamma f(\gamma,t).
\end{equation}
In the steady state, the flux $J$ is required to vanish identically, and for an angular velocity $\Omega_z$ of the form \eqref{eq:Omega_goles} the orientational distribution reads 
\begin{equation}
  \label{eq:stat_prop_dist}
  f(\gamma)
  =
  N^{-1}
  \te{exp}
  \left\{
    \frac{\Omega_1}{D_\te{r}}
    \cos\gamma
    +
    \frac{\Omega_2}{D_\te{r}}
    \cos^2\gamma
  \right\},
\end{equation}
with a normalization factor\footnote{
$
  N
  =
    \te{e}^{B-A}
    \left.
    \left[
      \te{e}^{2A}
      \mathcal D
      \left(
        \frac{A+2B}{2\sqrt{B}}
      \right)
      -
      \mathcal D
      \left(
        \frac{A-2B}{2\sqrt{B}}
      \right)      
    \right]
    \right / 
    \sqrt{B}$
with $A \equiv \Omega_1/\Dr$, $B \equiv \Omega_2/\Dr$, and 
$
\mathcal D(x)
\equiv
\te{e}^{-x^2}
\int_0^x \df y ~ \te{e}^{y^2}.
$ } $N$.

\subsection{Experimental Results}
\label{sec:experimental-results}

\begin{figure}	\includegraphics[width=1\columnwidth]{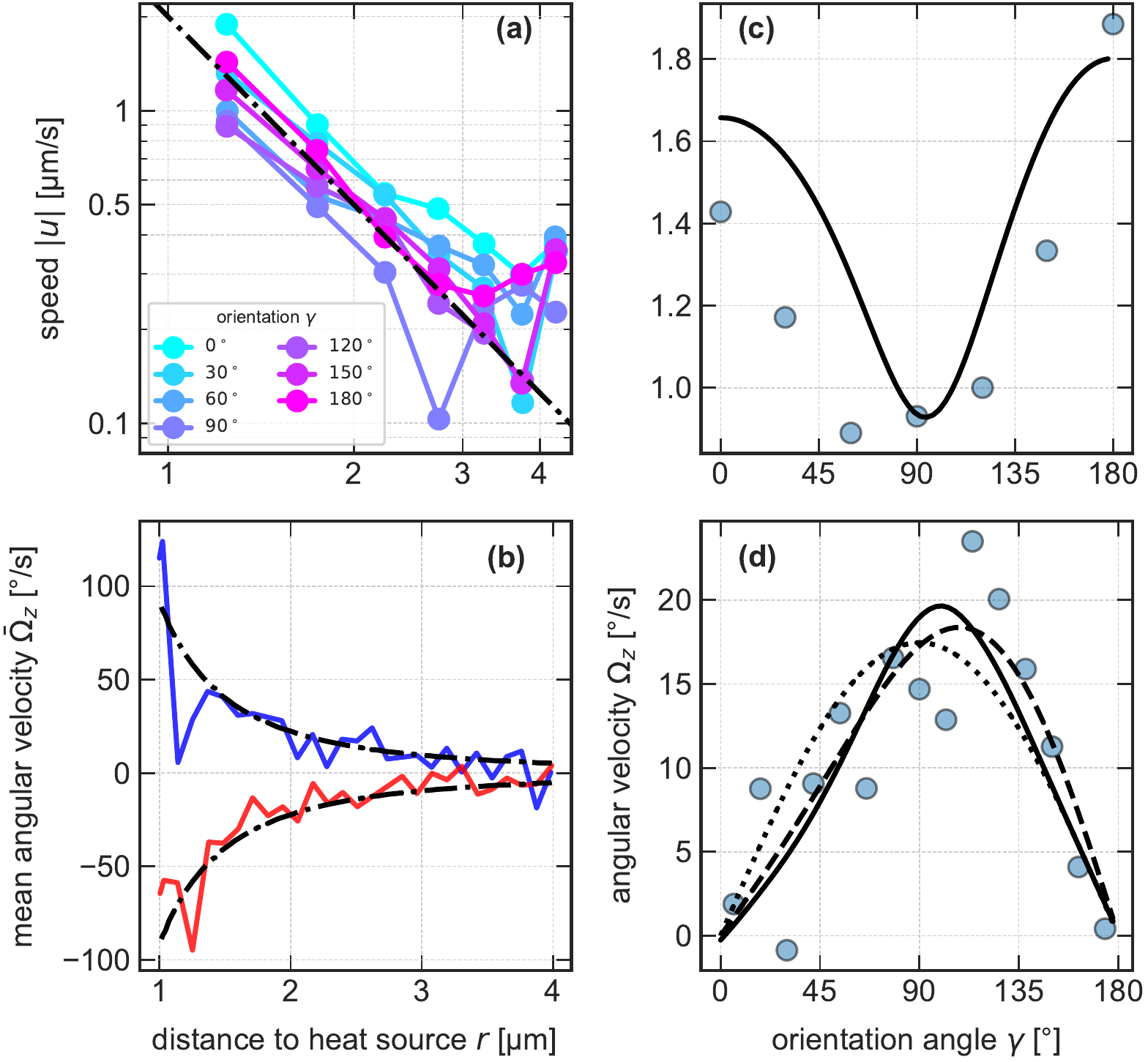}
  \caption{%
  \textbf{Distance- \textbf{(a,b)} and orientation- \textbf{(c,d)} dependence of translational and (mean) angular swim speed.}
  The Janus particle's swim speed $u$ \textbf{(a)} and mean angular speed $\bar \Omega_z$ (averaged over initial orientations $\gamma$) \textbf{(b)} both decay like $r^{-2}$ (dash-dotted line) in the distance $r$ from the heat source, as expected from Fourier's law of heat diffusion.
Upper and lower branch in panel \textbf{(b)} correspond to clockwise and counterclockwise rotation, respectively.
The orientational dependence of the  swim speed in  panel  \textbf{(c)}, measured at a distance $r = \SI{1.25}{\micro\metre}$ from the heat source, conforms with the theoretical fit
        \(
        \left(
          u_x^2+u_y^2
        \right)^{1/2}
        \) with $u_{x,y}$ obtained from Eqs.~\eqref{eq:theory_ux}, \eqref{eq:theory_uy} using numerically determined temperature profiles  (see Fig.~\ref{fig:surf_temp_incr}). 
 \textbf{(d)}. 
    To resolve the orientation-dependence of the angular velocity $\Omega_z$, data in the interval $r = 1$-$\SI{4}{\micro\metre}$ was pooled. 
    The theoretical fit (solid curve) was obtained from Eq.~\eqref{eq:theory_omega}, again using the numerically determined temperature profiles.   The least-square fits in c) and d) yield $\mups = \SI{2.88}{\micro\metre^2 \per \second \kelvin}$ and $\muau = \SI{1.82}{\micro\metre^2 \per \second \kelvin}$ for the mobilities.
    The alternative fits shown in panel d) follow from Eq.~\eqref{eq:Omega_goles} with $\Omega_{1,2}$ as independent fit parameters (dashed) and  $\Omega_z = \Omega_0 \sin\gamma$ \cite{Bickel2014Polarization}, with $\Omega_0$ as fit parameter (dotted), and yield $\Omega_0 = \SI{17.4}{\degree/\second}$, $\Omega_1=  \SI{17.2}{\degree/\second}$, $\Omega_2= \SI{-3.25}{\degree/\second}$. 
  }
	\label{fig_exp_results}
\end{figure}
Figure~\ref{fig_exp_results} displays the experimental results for the magnitude of the phoretic propulsion speed $u(\gamma,r)$ as a function of the distance $r$ from and  orientation $\gamma$ to the heat source. The speed $u$ decays with the squared reciprocal distance, as expected for an external temperature gradient $\nabla T\propto 1/r^2$ consistent with Fourier's law.  The maximum speed is
\(
u
=
\SI{2}{\micro\metre / \second}
\)
at a distance of $r = \SI{1.25}{\micro\metre}$.  Closer to the heat source, tracking errors limit the acquisition of reliable data. The experiments also provide direct evidence for a thermophoretic rotational motion of the Janus particle. According to Eq.~\eqref{eq:theory_omega}, the boundary temperatures as well as the phoretic mobility coefficients must therefore differ between the gold and polystyrene parts of the particle. Figure \ref{fig_exp_results} (b) shows the mean angular velocity $\bar \Omega$ for clockwise ($+$) and counter-clockwise ($-$) rotation, with the mean over positive and negative values of the initial orientation $\gamma$ taken separately.  The angular velocity also decays with the squared reciprocal distance from the heat source from  
\(
\bar \Omega = \SI{100}{\degree / \second}
\)
at short distances.

The translational and rotational speeds depend on the orientation $\gamma$ to the heat source, due to the Janus-faced particle surface and its heterogeneous mobility coefficients $\mu$ and thermal conductivities $\kappa$. We have therefore also analyzed the  particle's motion as a function of the initial orientation $\gamma$. The experimental results are plotted in Figure \ref{fig_exp_results} (c) and (d).
For the translational phoretic speed $u$ we observe a clear minimum between $\gamma=\ang{50}$ and $\gamma=\ang{135}$. 
Local maxima are observed when the polymer side is facing the heat source ($\gamma=\ang{180}$) or pointing away from it ($\gamma=\ang{0}$). That the latter orientation displays a smaller speed suggests that the polymer side yields the major contribution. 

\begin{figure}
  \centering
\includegraphics[width=1\columnwidth]{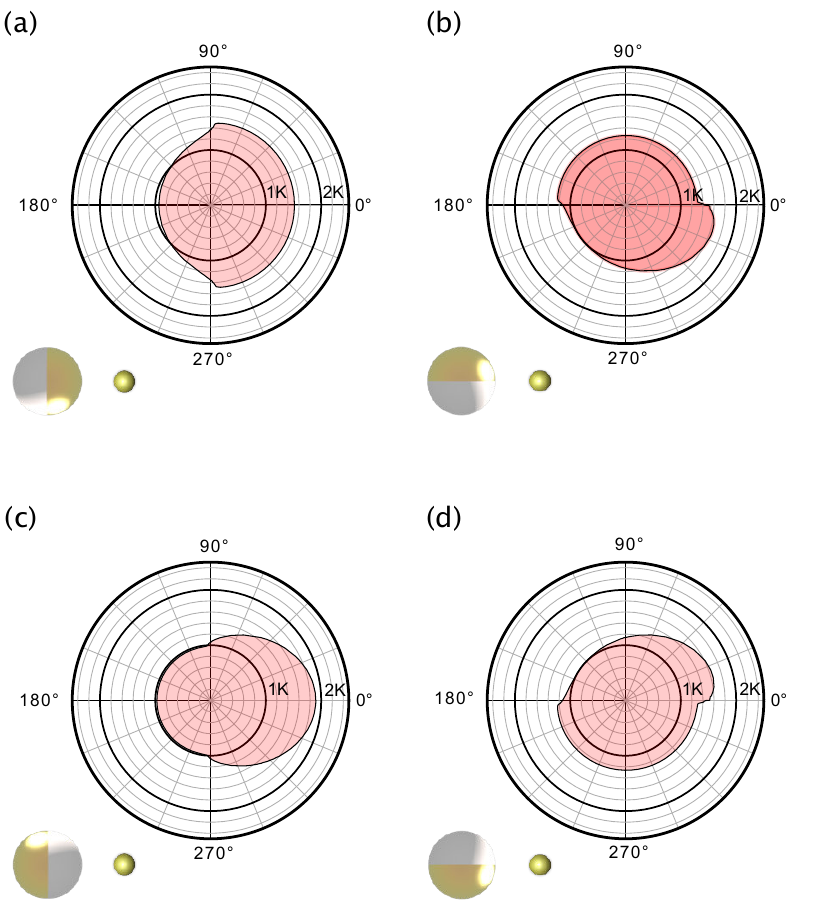}
  \caption{%
    \textbf{Numerically determined azimuthal temperature variations}. The temperature increments $\langle T \rangle_\theta(\phi) - T_0$ (in Kelvin) along the Janus particle circumference but averaged over the normal angle, are depicted  for 4 different orientations $\gamma$ and a fixed distance  $r=\SI{1.25}{\micro\metre}$ between the Janus particle and the source: (a) $\gamma=\ang{0}$, (b) $\gamma=\ang{90}$, (c) $\gamma=\ang{180}$, (d) $\gamma=\ang{270}$.}
  \label{fig:surf_temp_incr}
\end{figure}

 In spite of averaging $\Omega_z$ over the measured distance range ($\SI{1}{\micro\metre}$--$\SI{4}{\micro\metre}$) the $\gamma$-dependent angular velocity exhibits some residual scatter. It is still seen to vanish for $\gamma=\ang{0}$ and $\gamma=\ang{180}$ [Fig.~\ref{fig_exp_results} (d)], in line with the expected symmetry of the temperature field around the axis of the Janus particle. 
At $\gamma \approx \ang{90}$, we observe a maximum angular and minimum translational speed.

To compare the experimental results to our theoretical expectations \eqref{eq:theory_ux}--\eqref{eq:Omega_goles}, we require further information on the angular dependence of the temperature at the surface of the Janus particle. For this purpose, we numerically solved the complex heat conduction problem with a commercial PDE solver \cite{comsol} (\ref{sec:temp_comsol}). The  obtained profiles of the mean temperature increment $\av T_\theta(\phi) - T_0$ along the circumference of the Janus particle are displayed in Fig.~\ref{fig:surf_temp_incr}. They reveal that the largest temperature difference between the gold (au) and polystyrene (ps) side is attained when the polymer is facing the heat source, confirming the experimental trend. They also exhibit unequal mean boundary temperatures 
\(
\av{T}_{\theta}|_{\phi=3\pi/2}
\)
and
\(
\av{T}_{\theta}|_{\phi=\pi/2}
\)\,, 
as required by Eq. \eqref{eq:theory_omega} for angular motion.

The experimental results on the translational and the angular velocity as a function of the orientation angle $\gamma$ can be compared to the theoretical predictions \eqref{eq:theory_ux}--\eqref{eq:theory_omega} while using the numerically calculated surface temperature profiles to obtain estimates for the phoretic mobility coefficients pertaining to the different surface regions of the Janus particle. 
A least-square fit of the theoretical prediction \eqref{eq:theory_omega} for the angular velocity $\Omega_z$ yields our best estimate for  $\mups - \muau$.
Inserting it into Eqs.~\eqref{eq:theory_ux} and \eqref{eq:theory_uy} for the translational velocity components, another least-square fit for the phoretic speed $(u_x^2+u_y^2)^{1/2}$  eventually yields the optimum values
\(
\mups
=
\SI{2.88}{\micro\metre^2 \per \second \kelvin}
\)
and
\(
\muau
=
\SI{1.82}{\micro\metre^2 \per \second \kelvin}
\)
for the phoretic mobilities. The theoretical fits are shown in Fig.~\ref{fig_exp_results} (c,d) as solid lines, while the dashed line is a fit of Eq.~\eqref{eq:Omega_goles} with $\Omega_{1,2}$ as independent fit parameters. It nicely reproduces the experimental data.
In contrast, assuming the rotational speed to be of the form $\Omega_z \propto \sin\theta$ \cite{Bickel2014Polarization} (dotted), as for  homogeneous heat conductivity, misses the experimentally observed asymmetry.

\begin{figure}
  \centering
  \includegraphics[width=\columnwidth]{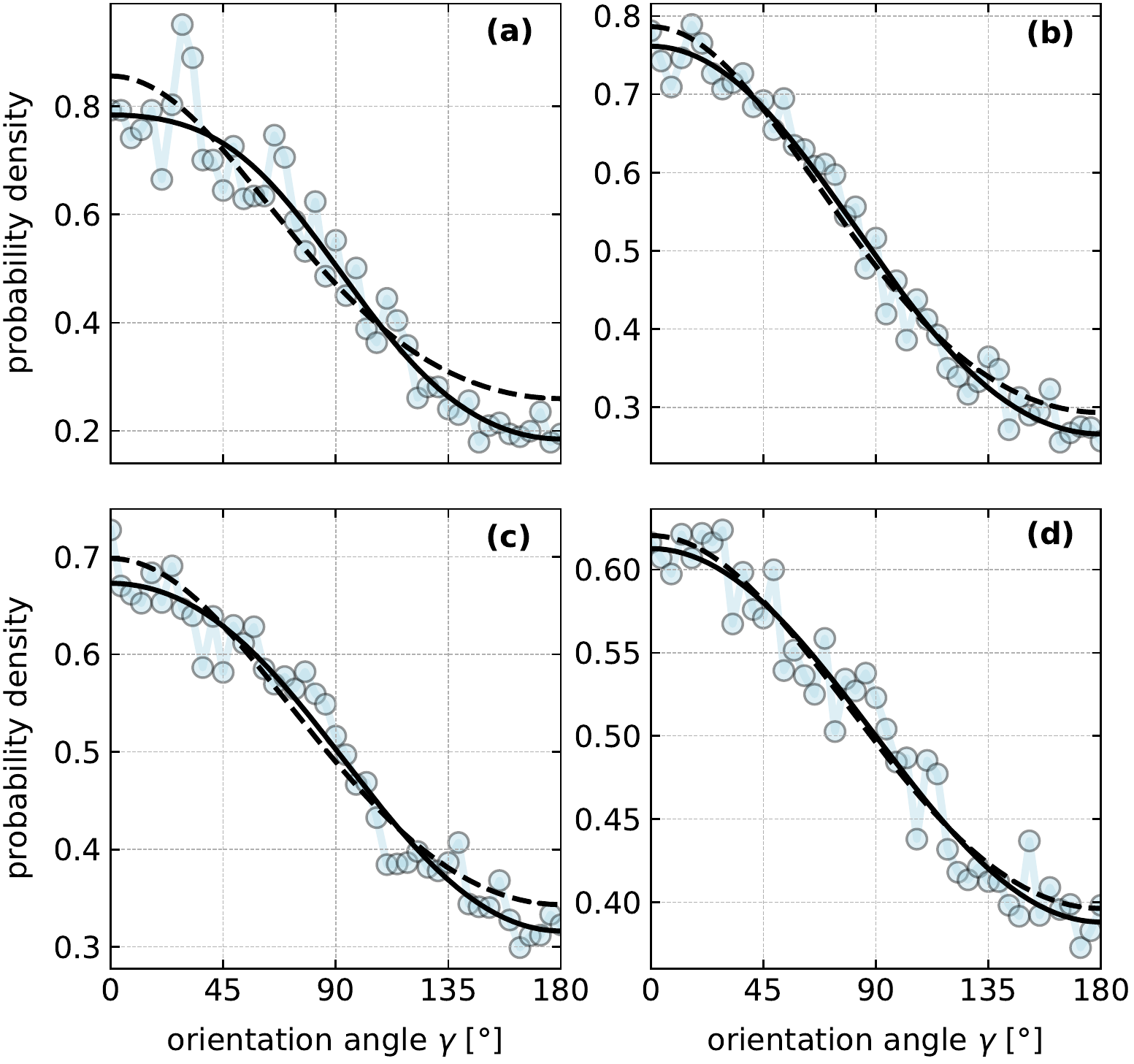}
  \caption{%
    \textbf{Probability density to find the Janus sphere pointing at an angle $\gamma$ to the heat source.} The panels show data (symbols) measured for various distances $r$ between particle and heat source: \textbf{(a)} \SI{1.1}{\micro\meter}, \textbf{(b)} \SI{1.7}{\micro\meter}, \textbf{(c)} \SI{2.3}{\micro\meter}, \textbf{(d)} \SI{2.8}{\micro\meter}. The solid  lines are best fits by Eq.~\eqref{eq:stat_prop_dist}, with free fit parameters  $\Omega_{1,2}/D_\te{r}$.
    The dashed lines are fits by $f \propto \te{exp}[(\Omega_0/D_\te{r})\cos\gamma]$ \cite{Bickel2014Polarization} for an angular speed profile  $\Omega_z = \Omega_0 \sin\gamma$ with $\Omega_0/D_\te{r}$ as a free fit parameter. These fits yield for $\Omega_0/\Dr$, $\Omega_1/\Dr$, $\Omega_2/\Dr$ the values  \textbf{(a)} 0.596, 0.722, -0.284, \textbf{(b)} 0.493, 0.526, -0.091, \textbf{(c)} 0.355, 0.377, -0.0855, \textbf{(d)} 0.224, 0.228, -0.0248, respectively.
  }
  \label{fig:probabilities}
\end{figure}

Besides these dynamical properties, we also assessed the stationary distribution of the Janus particle's orientation relative to the heat source, at various distances. Figure \ref{fig:probabilities} verifies that the particle aligns with the external temperature gradient. 
In accordance with the positive angular velocities observed for $0 < \gamma < \ang{180}$ in Fig.~\ref{fig_exp_results} (d), we measure a significantly higher probability to find the particle's gold cap pointing towards the heat source than away from it.

\subsection{Discussion}
\label{sec:discussion}
The motion of a colloidal particle in an external temperature gradient is determined by the thermo-osmotic surface flows \cite{bregulla2016Thermoosmosis} induced by the temperature gradients along the particle's surface  via its physio-chemical interactions with the solvent. Knowing both the temperature profile and interfacial interaction characteristics should thus allow the behavior of our Janus particle in an external temperature field to be explained. Note, however, that the heterogeneous material properties of the Janus particle matter in two respects. First, if the hemispheres do not have the same heat conductivities, this will distort the temperature profile in the surrounding fluid in an unsymmetric, orientation-dependent manner. Secondly, their generally unequal thermo-osmotic mobility coefficients $\mu$ will translate the resulting surface temperature gradients differently into phoretic motion.   
The numerically determined temperature profiles for our Janus particle, shown Fig.~\ref{fig:surf_temp_incr}, reveal that the presence of the Janus particle indeed distorts the external field significantly, and that the difference between the heat conductivities of the two hemispheres matters. The large thermal conductivity of gold creates an almost isothermal temperature profile on the gold cap (even if the thin film conductivity is somewhat lower than the bulk thermal conductivity). The resulting temperature distribution is for some orientations $\gamma$ reminiscent of the temperature distribution on the surface of a self-propelled Janus particle. In the latter case, the metal cap itself is the major light absorber and thus the heat source creating the surrounding temperature gradient. In our case the gradient is primarily caused by the external heat source, but modulated by the presence of the Janus sphere.  Unless the particle's symmetry axis is perfectly aligned with the heat source [Fig.~\ref{fig:surf_temp_incr} (a),(c)], the mean temperature profile is generally asymmetric along the particle's circumference. Such asymmetric distortions of the temperature field were not considered in previous theoretical studies \cite{Bickel2014Polarization} but matter for the proper interpretion of Eqs.~\eqref{eq:theory_ux}--\eqref{eq:theory_omega} for the particle's linear and angular velocities.

Equation \eqref{eq:theory_uy} yields the transverse thermophoretic velocity, $u_y$, of the particle, \ie, the velocity perpendicular to its symmetry axis.  Assuming a constant temperature on the gold hemisphere, the only contribution for the transverse velocity $u_y$ results from the temperature gradients along the polystyrene side --- due to the $\sin\phi$ term in Eq.~\eqref{eq:theory_uy} --- and $u_y$ is determined by the mobility coefficient $\mups$.
The velocity component $u_x$ along the particle's symmetry axis contains two terms according to Eq.~\eqref{eq:theory_ux}. The first term yields a propulsion along the symmetry axis to which both hemispheres contribute according to the $\cos\phi$ term. It tends to suppress the details at the au--ps interface, where the temperature gradients are typically most pronounced. Hence, the temperature profile in the vicinity of the particle poles and the corresponding mobilities largely determine the first term in Eq.~\eqref{eq:theory_ux}. The second term, which  only depends on the boundary values of the (weighted) mean temperature at the au-ps interface and the mobility step
\(
\mups - \muau
\), 
is of opposite sign and thus reduces the total propulsion velocity. (It disappears if $\mups \approx \muau$.)

\begin{figure}	\includegraphics[width=1\columnwidth]{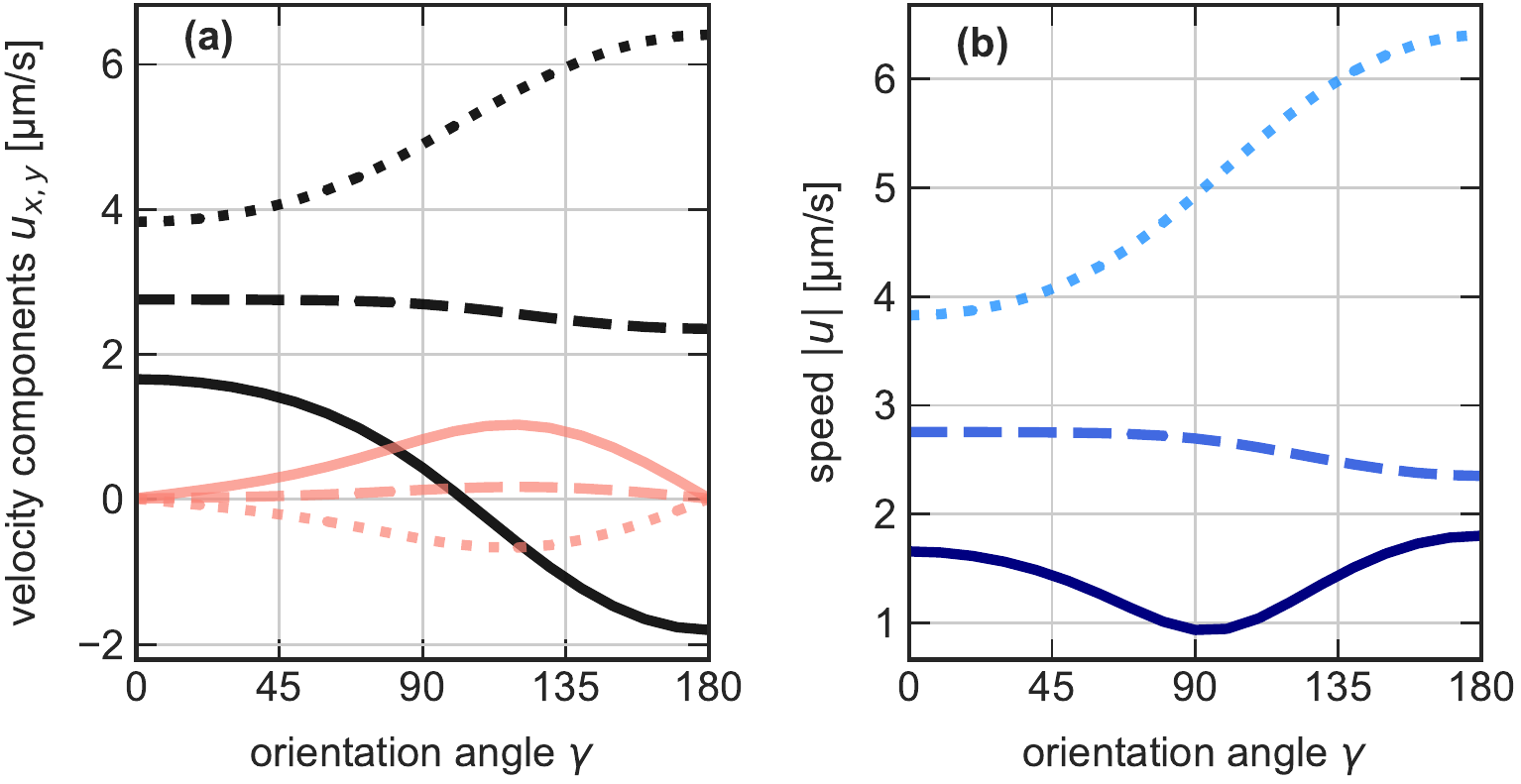}
  \caption{%
    \textbf{Orientation dependence of the phoretic propulsion} \textbf{(a)}  Longitudinal and transverse velocity components $u_x$ and $u_y$ ($u_y$  starting at the origin) from Eqs.~\eqref{eq:theory_ux} and \eqref{eq:theory_uy}, respectively.   For the phoretic mobilities, three  sets of values are considered:  those obtained from the fits in Fig.~\ref{fig_exp_results} (b,d), namely
    $\mups=\SI{2.88}{\micro\metre^2 \per \second\kelvin}$  and $\muau=\SI{1.82}{\micro\metre^2 \per \second\kelvin}$ (solid lines); $\mups=\SI{0.5}{\micro\metre^2 \per \second\kelvin}$ and $\muau=\SI{-0.55}{\micro\metre^2 \per \second\kelvin}$ (dashed); $\mups=\SI{-1.82}{\micro\metre^2 \per \second\kelvin}$ and $\muau=\SI{-2.88}{\micro\metre^2 \per \second\kelvin}$ (dotted);
    \textbf{(b)} the corresponding total propulsion speeds
    \(
    \left(
      u_x^2+u_y^2
    \right)^{-1/2}
    \).
  }
  \label{fig:plausibility_mu}
\end{figure}

Figure~\ref{fig:plausibility_mu} illustrates the  orientation dependence of the phoretic velocity compoents  $u_x$ and $u_y$  obtained from Eqs.~\eqref{eq:theory_ux} and \eqref{eq:theory_uy}. The longitudinal component $u_x$ (along the particle's symmetry axis) is positive or negative depending on whether the ps-hemisphere faces away from or towards the heat source. Its smooth sign change at $\gamma \approx \ang{90}$ simply reflects the fact that the interaction is  overall repulsive. Notice, however, that the higher thermal conductivity of the gold cap creates a surface temperature contribution mimicking that for an optically heated Janus swimmer. The ensuing (self-) propulsion along the $x$ direction shifts the zero crossing slightly from $\ang{90}$. This thermophoretic "swimmer-contribution" to the propulsion is not generally parallel to the direction of the external temperature gradient, unless it is perfectly aligned to the heat source, thereby causing subtle deviations from predictions for particles  with isotropic heat conductivity \cite{Bickel2014Polarization}. 
The transverse velocity component $u_y$ naturally vanishes if the particle axis is aligned or anti-aligned with the heat source ($\gamma = \ang{0}$ and $ \gamma = \ang{180}$). It attains a maximum at $\gamma \approx \ang{120}$, when the polystyrene hemisphere is oriented somewhat towards the heat source, which allows for the maximum lateral surface temperature gradients.
For the same reason, the maximum propulsion speed $u$ is attained for $\gamma = \ang{180}$ (ps-side facing the heat source) and only a lesser  local maximum is seen at $\gamma = \ang{0}$ (au-side facing the heat source), in Fig.~\ref{fig_exp_results} (c).

From our fits in Fig.~\ref{fig_exp_results} we obtained $\mups > \muau > 0$. The first condition  ensures the correct sign for the angular velocity according to Eq.~\eqref{eq:theory_omega} and Fig.~\ref{fig_exp_results}(d), and is in agreement with previous findings for thermo-osmotic interfacial flows \cite{bregulla2016Thermoosmosis}. The step in the phoretic mobility at the particle equator determines the magnitude and sign of the angular velocity \cite{Bickel2014Polarization}. While different absolute values can lead to the same step height $\mups-\muau$,  the $\gamma-$dependence of the translational velocity also constraints these absolute values. This is illustrated by the dashed and dotted lines in Fig.~\ref{fig:plausibility_mu}, representing other combinations of phoretic mobilities, including negative signs ($\mups >0,~\muau <0$ or $\muau<0<\mups$). Such choices would result in a quantitative and qualitative mismatch between theory and data. They also serve to demonstrate that the  motion of the particle is very sensitive to these values, for a given temperature profile. 

According to Eq.~\eqref{eq:theory_omega}, the angular velocity component $\Omega_z$ only depends on the equatorial interfacial values  at $\phi=\pi/2$ and $3\pi/2$ of the average temperature $\av{T}_\theta$ and the jump in the mobility coefficients. In other words, the details of the temperature profile on both sides of the particle are irrelevant for the rotational motion as long as the two boundary temperatures and the two mobility coefficients differ appreciably, but rotational motion will cease if either pair coincides. Hence, irrespective of the negligible temperature gradient on the gold side, the rotational velocity is sensitive to the thermo-osmotic mobility coefficient $\muau$, which can thus confidently be inferred from the measurement.
Compared to the substantial thermal-conductivity contrast, the role of mass anisotropy, which can lead to similar polarization effects \cite{Olarte_Plata_2018thermophorTorque,Gittus2019thermalOrient,Olarte_Plata2020orientJPmassAnisotrop}, plays presumably a negligible role in our experiments, as the thin gold cap makes the Janus particle only slightly bottom heavy.

\section{Conclusions}
To summarize, we have investigated the interaction of a single gold-capped Janus particle with the inhomogeneous temperature field emanating from an immobilized gold nanoparticle. The setup allows for a precise and well-controlled study of thermophoretic interparticle interactions that dominate in dilute suspensions of thermophoretic microswimmers.  
To our knowledege, this is the first time, the repulsion of the Janus particle from the heat source and its thermophoretically induced angular velocity have quantitatively been measured. 
An interesting consequence of the induced angular motion is an emerging  polarization of the Janus particle in the thermal field, which should generalize to any type of Janus swimmer in a motility gradient.  In our case, it means that the metal cap preferentially points towards the heat source. 

In combination with numerically determined surface temperature profiles for various particle-heat source orientations, the standard hydrodynamic model for colloidal phoretic motion was found to nicely reproduce our experimental data. Theory and observation corroborate that the rotational motion hinges on two necessary conditions: (i) the phoretic mobilities of the Janus hemispheres must be distinct and (ii) the values of the driving field (in our cases the temperature) must differ accross the equator --- irrespective of its behavior in between.
In return, we could therefore infer the phoretic mobilities  from the observed rotational and translational motion in an external field gradient. We found them to be positive for both polystyrene and gold. 

As an interesting detail, we found that the distinct heat conductivities moreover break the naively expected symmetry of the particle's translational and rotational speeds as a function of the orientation, and, accordingly, of the resulting polarization of the Janus sphere with respect to the heat source. The observed asymmetries are quantitatively explained by the high heat conductivity of gold, which renders the metal cap virtually isothermal. This induces a robust translational motion that mimicks the self-propulsion of a  Janus swimmer in its self-generated temperature gradient, along it symmetry axis.
Since phoresis generally involves gradients in some (typically long-ranged) thermodynamic fields, our principal results should also apply to similar setups involving other types of phoretic mechanisms.

\vspace{1cm}

\noindent
\textbf{Acknowledgements: }
We acknowledge financial support by the Deutsche Forschungsgemeinschaft via SPP 1726 "Microswimmers" (KR 3381/6-1, KR 3381/6-2 and CI 33/16-1, CI 33/16-2).

\vspace{1cm}

\noindent
\section*{Author Contribution Statement}
AB conducted the experiments and analyzed the data. FC carried out the numerical calculations. SA carried out analytical calculations. KK, FC and SA drafted the manuscript.

\appendix

\section{Preparation of the Janus particles}
\label{sec:prep-janus-part}

\begin{figure}
  \centering
	\includegraphics[width=\columnwidth]{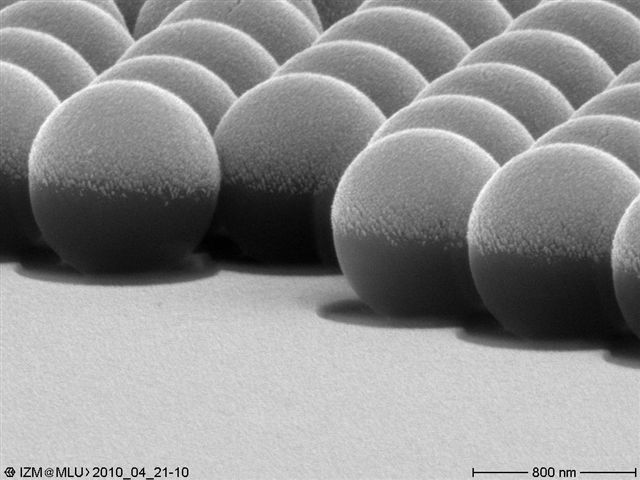}
  \caption[]{Raster electron microscopy image of the prepared Janus Particles.}
\label{fig:REM}
\end{figure}

The Janus particles have been prepared on standard microscopy glass cover slips, which have been treated in an oxygen plasma.  
A solution of polystyrene beads ($R = \SI{0.25}{\micro\metre}$; Microparticles GmbH) is deposited on these cover slips in a spin coater at $\SI{8000}{rpm}$.  
The particle concentration of the bead solution has been adjusted such that the particles do not form a closed packed monolayer but settle as rather isolated particles.  
This reduces the number of aggregates formed during the gold layer depostion.  
The samples have been further covered with a $\SI{5}{\nano\metre}$ chromium and a $\SI{50}{\nano\metre}$ gold film by evaporation in a vacuum chamber.  
The chromium layer has been added to make sure that the gold layer adheres to the glass slide when removing the Janus particles from the glass substrate by sonification.  
Fig.~\ref{fig:REM} displays a REM image of the prepared particles.

\section{Sample preparation}

The samples consist of two glass cover slips, which were rinsed with acetone, ethanol and deionized water, and treated with an oxygen plasma. They have been further coated with Pluronic F-127 (Sigma Aldrich) in a $\SI{5}{\percent}$ aqueous solution for a few hours. The Pluronic is adsorbed to the glass surface and residual Pluronic has been removed by rinsing the coated slides with deionized  water. A mixture of Janus particles and $R=\SI{125}{\nano\metre}$ gold colloids (British Biocell) was then deposited between the two slides and sealed with polydimethylsiloxane (PDMS) to prevent evaporation of the solution. The typical thickness of the liquid layer between the glass slides has been adjusted to be on the order of the diameter of the Janus particle ($\approx \SI{1}{\micro\metre}$) to prevent motion in vertical direction.

\section{Experimental setup}

The experimental setup consists of 2 parts: the heating and the illumination part. For the heating part a common laser source at a wavelength of $\SI{532}{\nano\metre}$ was used. This beam was first enlarged by a beam expander to fully illuminate an acousto-optic-deflector (AOD). This AOD is utilized to freely steer the focused beam within the sample. The optical path is arranged such that the beam waist is approximately $\SI{500}{\nano\metre}$. This beam is then focused by an oil immersion objective lens (Olympus 100x NA 0.5-1.3) into the sample.\\
The illumination of the sample is realized by an oil immersion dark field condenser (Olympus NA 1.2-1.4). The scattered white light is collected by the objective and imaged on the CCD-camera (Andor iXon). For the spatial position of the sample a piezo-scanner was used (Physik Instrumente, PI).

\begin{figure}
  \centering
 	\includegraphics[width=\columnwidth]{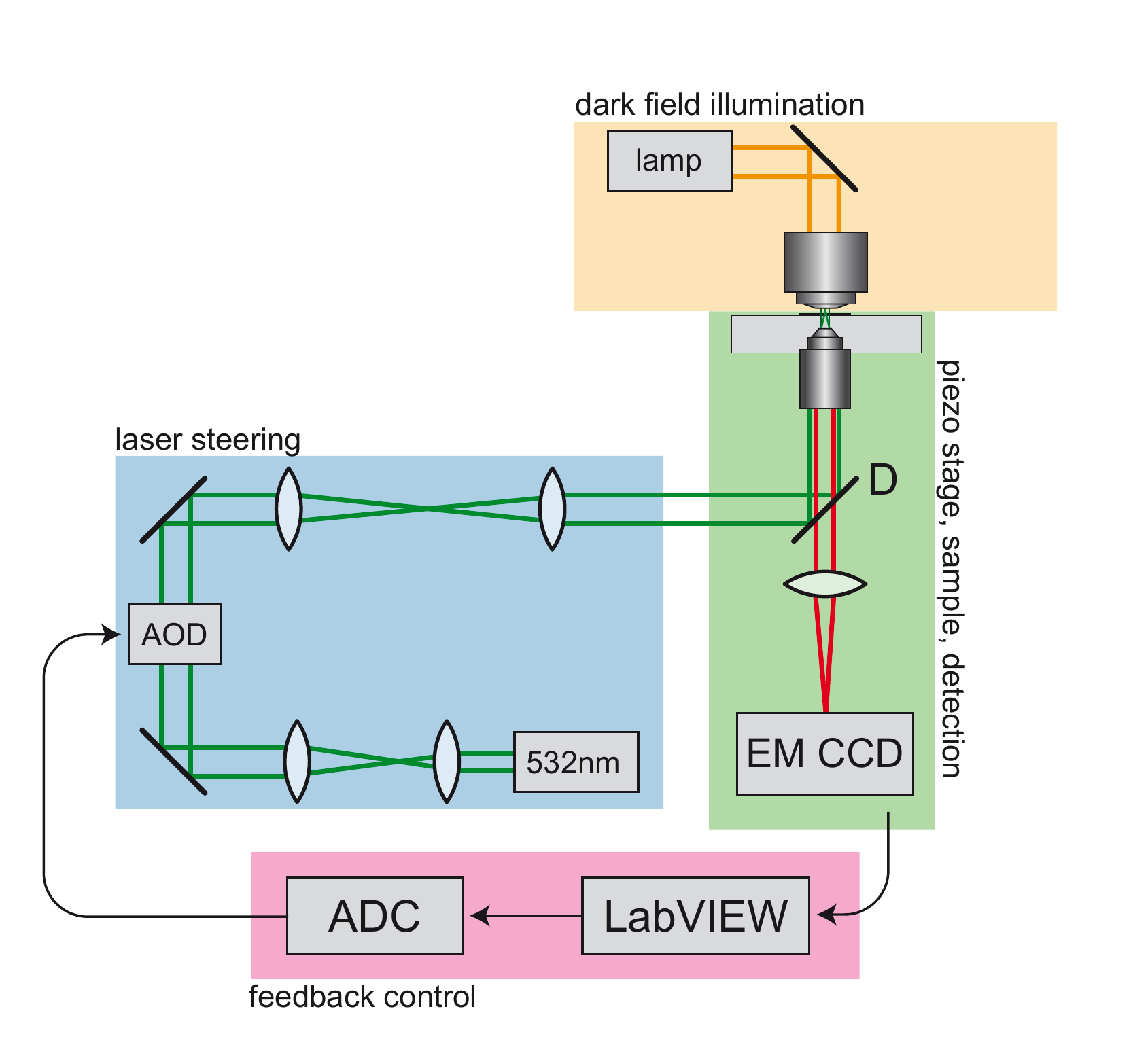}
  \caption[]{Experimental setup used for the experiments. See text for explanations and additional details.}
\label{fig:temp}
\end{figure}

\section{Particle tracking}
\label{sec:particle_tracking}
To determine the position of the Janus particles a binary image at a threshold above the background noise was taken.  The particle with the larger visual area was identified as  the Janus particle the smaller one as the gold heat source. The geometric centers of the visual images was identified with the particle position. For small distances ($<\SI{1}{\micro\metre}$) beween the Janus particle and the gold particle, the determination of the position of both particles fails. In this case the data is disregarded.\\
The image of he Janus particle is further analyzed with multiple binary images that are obtained by limiting the maximum image intensity to thresholds between $\SI{10}{\percent}$ and $\SI{80}{\percent}$. For each binary image the geometric center is determined. The $x-$ and $y-$coordinates of the geometric center are fitted with $f_{\rm{x}}=a_{\rm{x}}\,x+b_{\rm{x}}$ and $f_{\rm{y}}=a_{\rm{y}}\,x+b_{\rm{y}}$, respectively. From both fits, the in-plane orientation can be determined  by $\gamma=\arctan\left(a_{\rm{x}}/a_{\rm{y}}\right)$.

\section{Measurering the  temperature profile on  the surface of the heated gold nanoparticle}
\label{sec:meas-temp-prof}
\begin{figure}
  \centering
 	\includegraphics[width=\columnwidth]{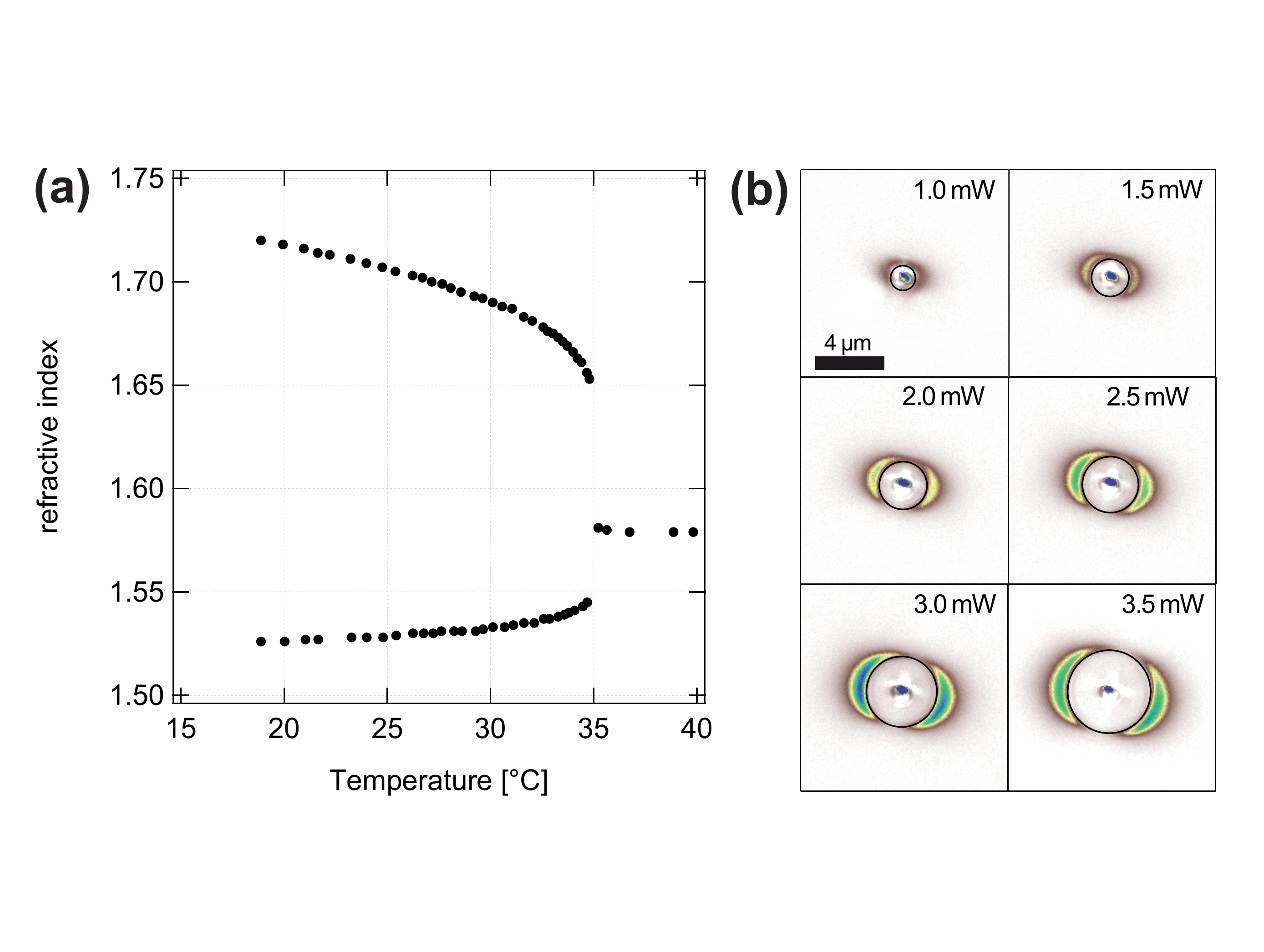}
  \caption[]{\textbf{(a)} The refractive index of 5CB in the nematic phase ($<\SI{35}{\celsius}$) and the isotropic phase ($>\SI{35}{\celsius}$)\cite{HornR.G.1978} \textbf{(b)} Example images of the growing isotropic bubble with heating power.  }
\label{fig:temp}
\end{figure}

For the estimate of the temperature increment $\Delta T_{\rm{Au}}$ at the surface of the gold nanoparticle heat source an additional experiment has been performed. In this experiment the solvent  was replaced by a liquid crystal (5CB). Its nematic-to-isotropic  melting transition upon heating beyond  $T_{\rm{ph}}=\SI{35}{\celsius}$ \cite{HornR.G.1978,Marinelli1998} was employed as a temperature sensor. The colloidal heat source generates a radial  temperature field 
\begin{equation}
T\left(r\right)=T_{\rm{0}}+\frac{\Delta T_{\rm{Au}}\left(r\right)R}{r}=T_0+\frac{P_{\rm{abs}}}{4\pi \kappa r}
\end{equation}
with $\kappa$ being the thermal conductivity of the medium, $P_{\rm{abs}}$ the absorbed power, proportional to the incident light power $P_{\rm{inc}}$, $T_{\rm{0}}=\SI{22}{\celsius}$ the ambient temperature, and $R=\SI{125}{\nano\metre}$ the radius of the gold colloid.\\
Whenever the temperature exceeds the phase transition temperature  $T\left(r\right)>T_{\rm{ph}}$ the nematic order melts. Since the molecular temperature field  $T\left(r\right)$ varies locally, the phase transition is confined to the vicinity of the heat source. Due to the radially symmetric shape of the temperature profile, an isotropic bubble forms around the gold colloid if $\Delta T_{\rm{Au}}+T_{\rm{0}}>T_{\rm{ph}}$. The size $r_{\rm{ph}}$ of the bubble scales linearly with $\Delta T_{\rm{Au}}$ and therefore with the heating power $P_{\rm{inc}}$:
\begin{equation}
r_{\rm{ph}}=\frac{\Delta T_{\rm{Au}}R}{\left(35^\circ-T_{\rm{0}}\right)}\sim\Delta T_{\rm{Au}}\sim P_{\rm{inc}}
\end{equation}

\begin{figure}
  \centering
 	\includegraphics[width=\columnwidth]{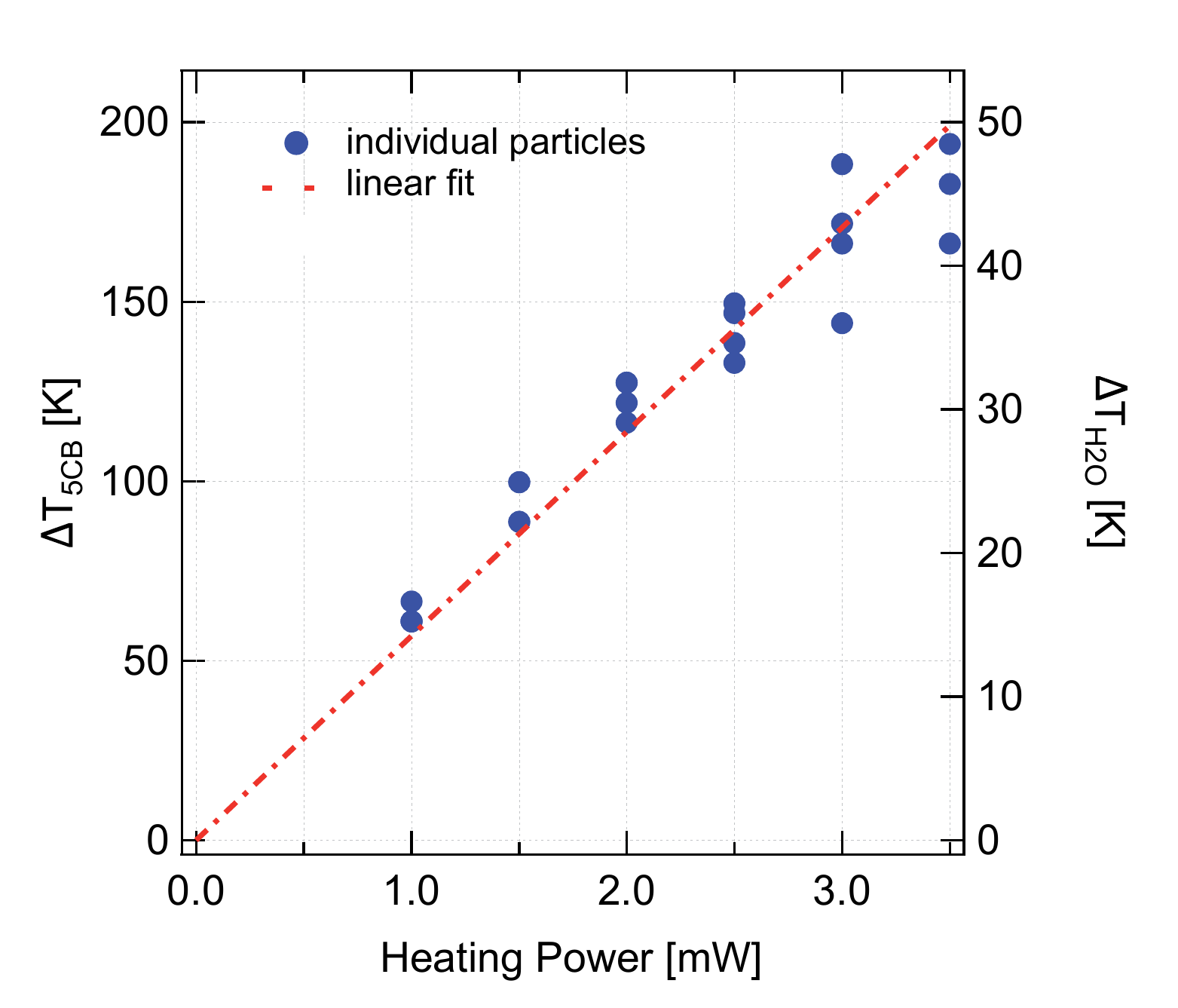}
  \caption[]{\textbf{Surface temperature increment  on the gold colloid} (relative to the ambient temperature of the solvent) for  a liquid-crystalline solvent  and water, respectively. Symbols represent the experiment. The dashed line is a linear fit to the data to extract the temperature increment per heating power. The values in water have been calculated from the known thermal conductivities of the liquid crystal and water. 
  } 
\label{fig:restemp}
\end{figure}

In the experiments, the size of the isotropic bubble $r_{\rm{ph}}$ as the function of the incident power $P_{\rm{inc}}$ is of interest. Its observation in the dark field setup,  (see Fig.~\ref{fig:temp}(a), \cite{HornR.G.1978}) exploits the refractive-index change upon melting.  Similar to a colloidal particle with a refractive index deviating from the surrounding material the molten bubble scatters the incident white light and appears as a bright ring in the dark field microscope. Example images are displayed in Fig.~\ref{fig:temp}(b) for different incident heating powers. The black circle indicates the estimated bubble size.
Its knowledge allows the surface temperature of the gold colloid in the liquid crystal to be estimate by:
\begin{equation}
\Delta T_{\rm{Au}}=\frac{\, r_{\rm{ph}}\left(35^\circ-T_{\rm{0}}\right)}{R}.
\end{equation}
Since the heat equation is linear, the estimate for the temperature increment in water is determined by its thermal conductivity  $\kappa_{\rm{H_2O}}=\SI{0.6}{\watt \per (\meter \kelvin)}$ relative to that of the isotropic liquid crystal, $\kappa=\SI{0.15}{\watt \per (\meter \kelvin)}$.
The result is displayed in Fig.~\ref{fig:restemp} where the approximate temperature increment in water is displayed in addition to the estimated temperature increment in 5CB in dependence on the heating power. From the linear fit a temperature increment per heating power of $\approx \SI{14}{\kelvin \per (\milli \watt)}$ can be obtained.

\section{Influence of the laser heating on the Janus particle}
\label{sec:infl-laser-heat}

\begin{figure}
  \centering
 	\includegraphics[width=\columnwidth]{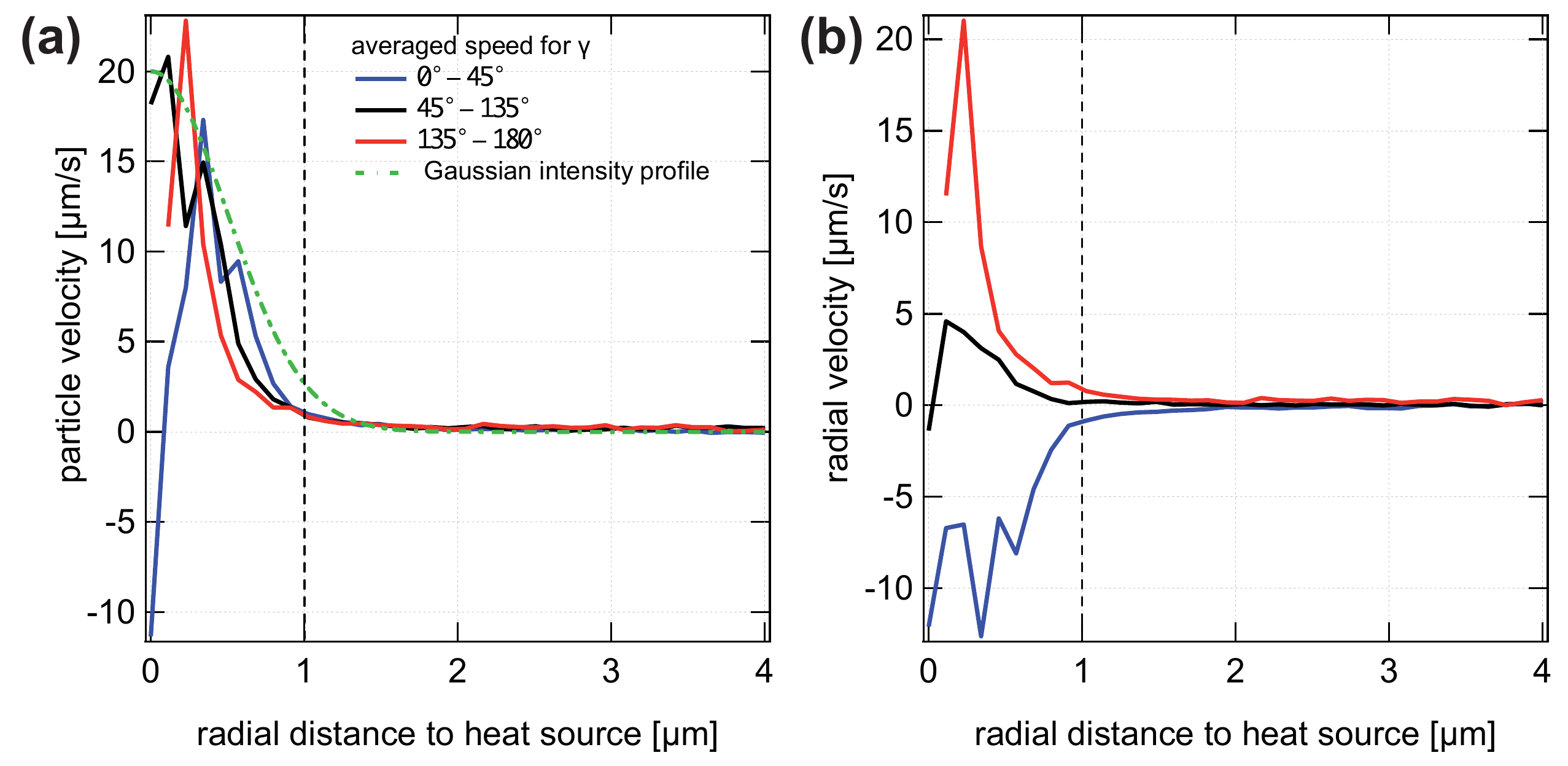}
  \caption[]{\textbf{(a)} Particle speed $u$ and \textbf{(b)} radial speed $v_{\rm{R}}$ in dependence on radial distance to the heat source for three orientations $\gamma$ of the particle relative to the heat source.  } 
\label{fig:noGold}
\end{figure}

The focused laser beam (beam-waist $\approx\SI{500}{\nano\metre}$) used for the heating of the immobile gold colloid may also heat the Janus particle directly. To quantify this effect, the experiment was repeated with and  without the immobile gold colloid with identical focus position. The influence of the laser beam can be estimated by calculating the particle velocity $u$ and the radial velocity $v_{\rm{R}}$.
The particle velocity is obtained by projecting the translational step $\Delta \mathbf{s}_i=\mathbf{r}_{i-1}-\mathbf{r}_{i}$ onto the particle orientation $\mathbf{n}^i_o$ and then performing the ensemble average $\mathbf{u}=\left\langle\Delta\mathbf{s}_i\cdot\mathbf{n}^i_o\right\rangle_i/\Delta t$ divided by the experimental timescale $\Delta t$ being the exposure time of the camera. Fig.\ \ref{fig:noGold} (a) displays the absolute value of $\mathbf{u}$ for 3 different orientations of the particle relative to the heat source $\phi$. The particle velocity is always positive as a result of the direct laser illumination, and quickly diminishes over a length scale comparable to the  laser beam width.  The radial velocity $v_{\rm{R}}=\left\langle \Delta \mathbf{s}_i\cdot\mathbf{e}^i_{\rm{R}}\right\rangle_i/\Delta t$ being the ensemble average of the scalar product of $\Delta\mathbf{s}_i$ and the unit vector in radial direction $\mathbf{e}^i_{\rm{R}}$ divided by the experimental timescale $\Delta t$ decays on similar length scales as the particle velocity.  Even though the influence of the direct laser heating on the Janus particle diminishes rather quickly with increasing distance, its influence  is still noticeable, and was therefore subtracted for the velocities presented in the main text.

\section{Derivation of the phoretic velocities}
\label{sec:derivation_theory}

 \subsection{The setup}
\label{sec:setup}
The considered setup of a Janus particle exposed to an external heat source, and conventions used in the following derivations, are summarized in Fig.~\ref{fig:method}.
The induced slip velocity $\vv v_{\rm s}$ at the surface of a Janus sphere of radius $a$ is given by [see Eq.~\eqref{eq:slip_velocity}]
\begin{equation}
  \label{eq:v_slip}
  \vv \vslip(\theta, \phi)
  =
  \mu(\theta, \phi)
  \vv \nabla_\parallel T(\theta, \phi),
\end{equation}
where the in-plane angle $\phi$ and normal angle $\theta$ are employed to parametrize the particle surface (rather than conventional polar coordinates adjusted to the particle symmetry). 
In the above equation, $\mu(\phi,\theta)$ is the thermophoretic mobility and
\begin{equation}
  \label{eq:nablaT}
  \vv \nabla_\parallel T
  \equiv
  \frac{\partial_\theta T}{a} \hat{\bm \theta}
  +
  \frac{\partial_\phi T}{a \sin \theta} \hat{\bm \phi}
\end{equation}
denotes the tangential part of the temperature gradient at the particle surface, expressed in terms of spherical coordinates. As they are constantly used in the following derivations, we note the corresponding unit vectors:
\begin{align}
  \label{eq:unit_vector_r}
  \hat{\bm r}
  &\equiv
  \begin{pmatrix}
    \cos\phi \sin\theta,
    \sin\phi \sin\theta,
    \cos\theta
  \end{pmatrix}^\top,
  \\
  \label{eq:unit_vector_theta}  
  \hat{\bm \theta}
  &\equiv
  \begin{pmatrix}
    \cos\phi \cos\theta,
    \sin\phi \cos\theta,
    -\sin\theta
  \end{pmatrix}^\top,
  \\
  \label{eq:unit_vector_phi}  
  \hat{\bm \phi}
  &\equiv
  \begin{pmatrix}
    -\sin\phi,
    \cos\phi,
    0
  \end{pmatrix}^\top.
\end{align}
The translational and rotational phoretic velocities, $\vv u$ and $\vv \Omega$, follow from $\vv \vslip$ as [see Eqs.~\eqref{eq:particle_velocity} and \eqref{eq:particle_angularvelocity}]
\begin{align}
  \label{eq:v_tr}
    \vv u
  &=
    - \frac{1}{|\mathcal S|}
    \oint\limits_{\mathcal S} \df S ~
    \vv \vslip
    =
      - \frac{1}{4 \pi}
      \int\limits_0^{2\pi} \df \phi
      \int\limits_0^\pi \df \theta ~ \sin \theta ~
      \vv \vslip,
  \\[0.5em]
  \nonumber
  \vv \Omega
    &=
      - \frac{3}{2 a}
      \frac{1}{|\mathcal S|}
      \oint\limits_{\mathcal S} \df
      \vv S \times \vv \vslip
  \\
  &=
    \label{eq:v_rot}
    - \frac{3}{8 \pi a}
    \int\limits_0^{2\pi} \df \phi
    \int\limits_0^\pi \df \theta ~ \sin \theta ~
    \hat{\vv r} \times \vv \vslip,
\end{align}
where $|\mathcal S| = 4 \pi a^2$ is the area of the particle surface $\mathcal S$.

It is experimentally observed that the particle preferentially aligns horizontally with the close-by cover slides. This observation enters our theory through the assumption that the swimmer rotates only about the $z$-axis, \ie, perpendicular to the observation plane. This implies that the swimmer also translates only in the $x$-$y$ plane.
Once the swimmer's $z$-axis remains invariant, the surface-temperature profile consequently always obeys the (approximate) symmetry 
\begin{equation}
  \label{eq.T_symm}
  T(\phi, \pi/2-\alpha)
  =
  T(\phi, \pi/2+\alpha),
  \qquad
  \alpha \in [0,\pi/2]
\end{equation}
 in the normal angle, in accord with the heterogeneous material composition of the Janus sphere. The local phoretic mobility $\mu$ may likewise be expressed as
\begin{equation}
  \label{eq:mu_decomp}
  \mu = \mu(\phi)
  \equiv
  \begin{cases}
    \mups \quad \te{for} \quad
    0 \leq \phi \leq \phipa
    \\
    \muau  \quad \te{for} \quad
    \phipa < \phi \leq \phiap
    \\
    \mups \quad \te{for} \quad
    \phiap < \phi < 2\pi
  \end{cases},
\end{equation}
where $\mups$ and $\muau$ are the constant phoretic mobilities corresponding to the polystyrene and gold part of the swimmer, respectively, and $\phipa$ and $\phiap$ denote the  angles pertaining to the equator between the distinct surface materials.

\subsection{Rotation}

We start with the term $\hat{\vv r} \times \vv \vslip$ inside the integral on the r.h.s.~of Eq.~\eqref{eq:v_rot}. Plugging in Eqs.~\eqref{eq:v_slip} and \eqref{eq:nablaT}, and using
\(
\hat{\vv r} \times \hat{\vv \theta}
=
\hat{\vv \phi}
\)
and
\(
\vv{\hat r} \times \boldsymbol{\hat{\phi}}
=
-\hat{\vv \theta}
\),
one obtains
\begin{align}
  \label{eq:cross_product}
  \hat{\vv r}
  \times
  \vv \vslip
  &=
    \frac{ \mu(\phi) }{ a }
    \left[
    (\partial_\theta T) \hat{\vv \phi}
    -
    \frac{\partial_\phi T}{\sin\theta}
    \hat{\vv \theta}
    \right]
  \\[0.5em]
  &=
  \frac{\mu(\phi)}{a}
  (\partial_\theta T)
  \begin{pmatrix}
    -\sin \phi
    \\
    \cos \phi
    \\
    0
  \end{pmatrix}
  \\
  &\quad+
  \frac{\mu(\phi)  (\partial_\phi T)}{a \sin\theta}
  \begin{pmatrix}
    -\cos\phi \cos\theta
    \\
    -\sin\phi \cos\theta
    \\
    \sin\theta
  \end{pmatrix}.
\end{align}
The symmetry relation \eqref{eq.T_symm}  implies
\begin{align}
  \label{eq:T_symm_deriv}
  \partial_\theta T(\phi, \pi/2-\alpha)
  &=
  -   \partial_\theta T(\phi, \pi/2+\alpha),
  \\[0.5em]
  \partial_\phi T(\phi, \pi/2-\alpha)
  &=
  \partial_\phi T(\phi, \pi/2+\alpha),
\end{align}
for $\alpha \in [0,\pi/2]$.
Hence, when calculating the surface average of Eq.~\eqref{eq:cross_product}, the $x$ and $y$-components vanish:
\begin{equation}
  \label{eq:Omega_xy_vanish}
  \int\limits_0^\pi \df \theta ~
  \sin\theta ~ \partial_\theta T
  =
  \int\limits_0^\pi \df \theta ~
  \sin\theta \cos\theta ~ \partial_\phi T
  =
  0.
\end{equation}
Via Eq.~\eqref{eq:v_rot}, the remaining $z$-component is given by
\begin{align}
  \nonumber
  \Omega_z
  &=
    -\frac{3}{8 \pi a}
    \int_0^{2\pi} \df \phi \,
    \mu(\phi)
    \int_0^{\pi} \df \theta \,
    \frac{\sin\theta}{a \sin\theta}
    \partial_\phi T(\theta, \phi)
    \sin\theta
    \\[0.5em]
  &=
    -\frac{3}{4 \pi a^2}
    \int_0^{2\pi} \df \phi \,
    \mu(\phi)
    \partial_\phi
    \av{T}(\phi),
\end{align}
where we introduced the mean ($\theta-$averaged) temperature $\av{T}_\theta(\phi)$ via
\begin{equation}
  \label{eq:T_averaged}
  \av{\bullet}_\theta(\phi)
  \equiv
  \frac{
    \int_0^\pi \df \theta ~
    \bullet(\phi, \theta) \sin\theta
  }{
    \int_0^\pi \df \theta ~
    \sin\theta
  }
  =
  \frac12
  \int_0^\pi \!\!\df \theta ~
  \bullet(\phi, \theta) \sin\theta.
\end{equation}
In contrast to Eqs.~\eqref{eq:theory_ux}--\eqref{eq:theory_omega} in the main text, we omit the subscript $\theta$ in the averaging notation $\langle \bullet \rangle$ throughout the rest of this section for the sake of brevity.
Using the mobility profile \eqref{eq:mu_decomp} and $2\pi$-periodicity of $\mu$ and $T$ in the angle $\phi$, the angular velocity simplifies to
\begin{align}
  \Omega_z
  &=
    -\frac{3}{4 \pi a^2}
    \Bigg(
    \mups
    \int\limits_0^{\phipa} \df \phi \,
    \partial_\phi
    \av{T}(\phi)
    \\
    &\qquad+
    \muau
    \int\limits_{\phipa}^{\phiap} \df \phi \,
    \partial_\phi
    \av{T}(\phi)
    +
    \mups
    \int\limits_{\phiap}^{2\pi} \df \phi \,
    \partial_\phi
    \av{T}(\phi)    
    \Bigg)
  \\[0.5em]
  &=
    -\frac{3}{4 \pi a^2}
    \big\{
    \mups
    \left[
    \av{T}(\phipa)
    -
    \av{T}(\phiap)    
    \right]
    \\
    &\qquad\qquad+
    \muau
    \left[
    \av{T}(\phiap)
    -
    \av{T}(\phipa)    
    \right]    
    \big\}
  \\[0.5em]
  &=
    \frac{3}{4 \pi a^2}
    (\mups -\muau)
    \left[
    \av{T}(\phiap)
    -
    \av{T}(\phipa)
    \right].
\end{align}
The final experession yields Eq.~\eqref{eq:theory_omega} upon identifying $\phipa = \pi/2$ and $\phiap = 3\pi/2$ for a half-coated Janus sphere.

\subsection{Translation}
Plugging Eq.~\eqref{eq:nablaT} in to Eq.~\eqref{eq:v_slip}, and using the expessions for the unit vectors \eqref{eq:unit_vector_r}--\eqref{eq:unit_vector_phi}, the local slip velocity at the particle surface reads
\begin{equation}
  \label{eq:v_slip_spherical}
  \vv \vslip
  =
  \frac{(\partial_\theta T)\mu(\phi)}{a} 
  \begin{pmatrix}
    \cos\phi \cos\theta
    \\
    \sin\phi \cos\theta
    \\
    -\sin\theta
  \end{pmatrix}
  +
  \frac{(\partial_\phi T)\mu(\phi)}{a\sin\theta}
  \begin{pmatrix}
    -\sin\phi
    \\
    \cos\phi
    \\
    0
  \end{pmatrix}.
\end{equation}
Calculating the surface average of Eq.~ \eqref{eq:v_slip_spherical}, one finds that its $z$-component vanishes, because
\begin{equation}
  \label{eq:Uz_vanish}
  \int\limits_0^\pi \df \theta ~
  \sin^2\theta ~ \partial_\theta T
  =
  0,
\end{equation}
by virtue of the symmetry relation \eqref{eq:T_symm_deriv}.
We now decompose the remaining $x$ and $y$ components of the translational velocity into
\(
\vv u^{(\theta)} + \vv u^{(\phi)}
\), corresponding to the contributions $\partial_\theta T$ and $\partial_\phi T$ of the temperature gradient \eqref{eq:nablaT}, respectively. We furthermore apply integration by parts to get rid of the temperature gradients and deal with the bare temperature profiles instead.

\subsubsection*{$\boldsymbol{\partial_\theta}$-part}
\label{sec:partial_theta-part}

Using Eq.~\eqref{eq:v_tr}, the $\theta$-derivative of the temperature gradient \eqref{eq:nablaT} gives
\begin{align}
  \nonumber
  \vv u^{(\theta)}
  &\equiv
    -
    \frac{1}{|\mathcal S|}
    \oint\limits_{\mathcal S}
    \df S ~
    \frac{\mu(\phi)}{a}(\partial_\theta T)
    \begin{pmatrix}
      \cos\phi \cos\theta
      \\
      \sin\phi \cos\theta
    \end{pmatrix}
  \\[0.5em]
  \label{eq:U_theta_full}  
  &=
    -\frac{1}{4 \pi a}
    \int\limits_0^{2\pi} \df \phi ~
    \mu
    \begin{pmatrix}
      \cos\phi
      \\
      \sin\phi
    \end{pmatrix}  
  \int\limits_0^{\pi} \df \theta ~
  \sin\theta \cos\theta ~ \partial_\theta T.
\end{align}
Applying integration by parts and using
\begin{equation*}
  \partial_\theta (\sin\theta \cos\theta)
  =
  \cos^2\theta - \sin^2\theta,
\end{equation*}
the $\theta$-integral in Eq.~\eqref{eq:U_theta_full} can be written as
\begin{align}
  \nonumber
  \int\limits_0^\pi \df \theta ~
  \sin\theta \cos\theta ~ \partial_\theta T
  &=
  \int\limits_0^\pi \df \theta ~
  \sin\theta ~
  \left(
    \sin\theta
    -
    \frac{\cos\theta}{\tan\theta}
    \right)T
  \\[0.5em]
  \label{eq:partInt_theta}
  &=
  2 \av{
    \left(
      \sin\theta
      -
      \frac{\cos\theta}{\tan\theta}
  \right)T},
\end{align}
with the $\theta$-average as defined in Eq.~\eqref{eq:T_averaged}. 
The $\theta$-contribution to the velocity thus reads
\begin{align}
  \label{eq:U_theta}
  \vv u^{(\theta)}
  =
  -\int\limits_0^{2\pi}
  \frac{\df \phi ~ \mu(\phi)}{2 \pi a }
    \av{
    \left(
      \sin\theta
      -
      \frac{\cos\theta}{\tan\theta}
  \right)T}(\phi)
  \begin{pmatrix}
    \cos\phi
    \\
    \sin\phi
  \end{pmatrix}.
\end{align}

\subsubsection*{$\boldsymbol{\partial_\phi}$-part}
\label{sec:boldsymb-part}

Analogously, the $\phi$-derivative of the temperature gradient \eqref{eq:nablaT} contributes
\begin{equation}
  \label{eq:U_phi}
  \vv u^{(\phi)}
  \equiv
  -\frac{1}{4 \pi a}
  \int\limits_0^{2\pi} \df \phi ~
  \mu(\phi)
 \begin{pmatrix}
    -\sin\phi
    \\
    \cos\phi
  \end{pmatrix}  
  \int\limits_0^{\pi} \df \theta ~
  (\partial_\phi T).
\end{equation}
The $\phi$-derivative appearing on the r.h.s.~of the above equation can be pulled out of the first integral. The remaining $\theta$-integration of the bare temperature profile can be expressed as
\begin{equation}
  \label{eq:T_tilde}
  \int\limits_0^{\pi} \df \theta ~
  T(\phi,\theta)
  =
  2\av{\frac{T}{\sin\theta}}
  \equiv
  2\Tint(\phi),
\end{equation}
with the $\theta$-average $\av{\bullet}$ as defined in Eq.~\eqref{eq:T_averaged}.
With the profile \eqref{eq:mu_decomp} of the local phoretic mobility $\mu(\phi)$, one finds
\begin{align}
  \vv u^{(\phi)}
  \nonumber
  &=
    \int_0^{2\pi}
    \frac{\df \phi ~ \mu(\phi)}{2 \pi a}
    \begin{pmatrix}
      \sin\phi
      \\
      -\cos\phi
    \end{pmatrix} ~
    \partial_\phi \Tint(\phi)
  \\[0.5em]
  \nonumber
  &=
    \frac{\mups}{2 \pi a}
    \int\limits_0^{\phipa} \df \phi \,
    \begin{pmatrix}
      \sin\phi
      \\
      -\cos\phi
    \end{pmatrix}~
    \partial_\phi
  \Tint
  \\
  \nonumber
    &\quad+
    \frac{\muau}{2 \pi a}
    \int\limits_{\phipa}^{\phiap} \df \phi \,
  \begin{pmatrix}
    \sin\phi
    \\
    -\cos\phi
  \end{pmatrix}~
    \partial_\phi
  \Tint
  \\
  \label{eq:u_phi_prior_part}
    &\quad+
    \frac{\mups}{2 \pi a}
    \int\limits_{\phiap}^{2\pi} \df \phi \,
  \begin{pmatrix}
    \sin\phi
    \\
    -\cos\phi
  \end{pmatrix}~    
    \partial_\phi
    \Tint
  \\[0.5em]
  \nonumber
  &=
    \frac{\mups}{2 \pi a}
    \left.
    \begin{pmatrix}
      \sin\phi
      \\
      -\cos\phi
    \end{pmatrix} ~
    \Tint
    \right|_{\phiap}^{\phipa}
    -
    \frac{\muau}{2 \pi a}
    \left.
  \begin{pmatrix}
    \sin\phi
    \\
    -\cos\phi
  \end{pmatrix} ~
    \Tint
  \right|_{\phiap}^{\phipa}
  \\
  \label{eq:u_phi_after_part}  
    &\quad-
      \int_0^{2\pi}
      \frac{\df \phi~\mu(\phi)}{2\pi a}
  \begin{pmatrix}
    \cos\phi
    \\
    \sin\phi
  \end{pmatrix}    ~
    \Tint
  \\[0.5em]
  \nonumber
  &=
    \frac{\mups - \muau}{2 \pi a} \times
  \\
  \nonumber
    &\quad\left[
    \begin{pmatrix}
      \sin(\phipa)
      \\
      -\cos(\phipa)
    \end{pmatrix}    
    \Tint(\phipa)
    -
    \begin{pmatrix}
      \sin(\phiap)
      \\
      -\cos(\phiap)
    \end{pmatrix}
    \Tint(\phiap)
  \right]
  \\
  \label{eq:u_phi}  
    &\quad-
    \int\limits_0^{2\pi}
      \frac{\df \phi~\mu(\phi)}{2\pi a}    
  \begin{pmatrix}
    \cos\phi
    \\
    \sin\phi
  \end{pmatrix}
    \Tint,
\end{align}
where we applied integration by parts from \eqref{eq:u_phi_prior_part} to \eqref{eq:u_phi_after_part}, and exploited $2\pi$-symmetry.

\subsubsection*{Combining both contributions}
\label{sec:comb-both-contr}

Adding the results \eqref{eq:U_theta} and \eqref{eq:u_phi} for $\vv u^{(\theta)}$ and $\vv u^{(\phi)}$, one finally arrives at
\begin{equation}
  \label{eq:U_final}
  \begin{aligned}
  \begin{pmatrix}
    u_x \\ u_y
  \end{pmatrix}
  &=
  \frac{\mups - \muau}{2 \pi a} \times
  \\
    &\quad\Bigg[
    \begin{pmatrix}
      \sin(\phipa)
      \\
      -\cos(\phipa)
    \end{pmatrix}
    \av{\frac{T}{\sin\theta}}(\phipa)
    \\
    &\qquad\qquad-
    \begin{pmatrix}
      \sin(\phiap)
      \\
      -\cos(\phiap)
    \end{pmatrix}      
    \av{\frac{T}{\sin\theta}}(\phiap)
  \Bigg]
  \\[0.5em]
    &\quad-
    2 \int\limits_0^{2\pi}
      \frac{\df \phi~\mu(\phi)}{\pi a}    
  \begin{pmatrix}
    \cos\phi
    \\
    \sin\phi
  \end{pmatrix}    
    \av{T \sin \theta}(\phi),
  \end{aligned}
\end{equation}
where we replaced $\Tint(\phi)$ by $\av{T/\sin\theta}$ [see Eq.~\eqref{eq:T_tilde}] and used the identity
\(
1 + \sin^2\theta - \cos^2\theta
=
2 \sin^2 \theta
\)
to arrive at the term $\av{T \sin\theta}$ appearing inside the integral in the last line of Eq.~\eqref{eq:U_final}. Setting $\phipa = \pi/2$ and $\phiap = 3\pi/2$ renders Eqs.~\eqref{eq:theory_ux} and \eqref{eq:theory_uy}.

\section{Finite-element simulation of the  temperature field}
\label{sec:temp_comsol}

\begin{figure}
  \centering
  \includegraphics[width=0.9\columnwidth]{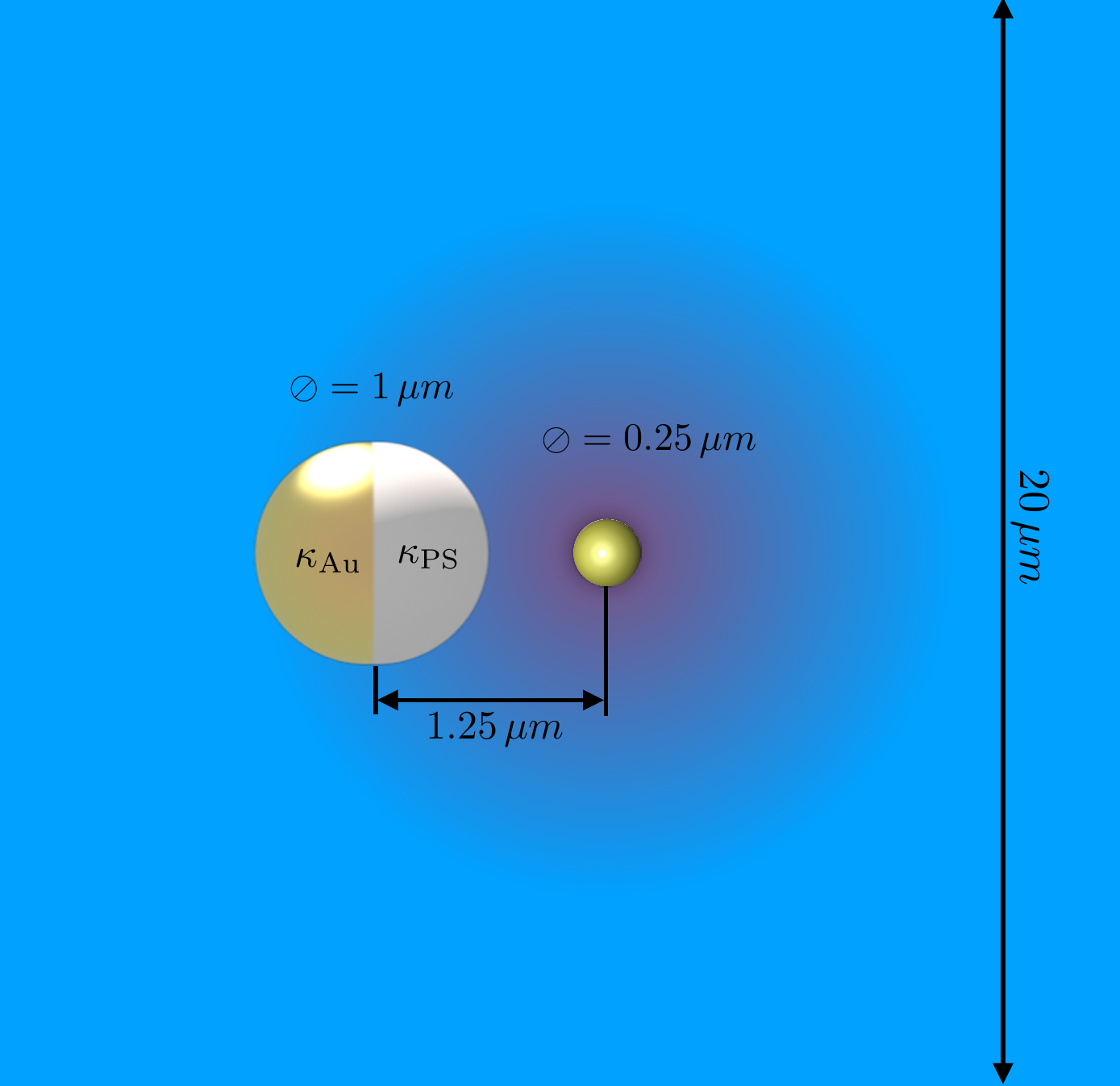}
  \caption{Geometry of the setup used for the numerical temperature calculations.}
  \label{fig:comsol_temp}
\end{figure}

To calculate the surface temperature of the Janus particle at different orientations, we use the COMSOL Multi-physics\textsuperscript{\tiny\textregistered} software \cite{comsol} to employ a finite-element solver for the considered heat conduction problem sketched in Fig.~\ref{fig:comsol_temp}.

The Janus particle is realized as a polystyrene particle of $\SI{1}{\micro\metre}$ diameter with a gold cap which is tapered to the edges and has a maximum thickness of $\SI{50}{\nano\metre}$. The heat source is a gold sphere of $\SI{250}{\nano\metre}$ diameter placed at $\SI{1.25}{\micro\metre}$ distance from the Janus particle center. Both particles are placed in a box of an edge length of $\SI{20}{\micro\metre}$.
The gold nanoparticle is heated with a heat source density of $\SI{1e15}{\watt\,\meter^{-3}}$. Other parameters used for the numerical calculations are listed in Table \ref{tab:simulation_parameters}.

\begin{table}[h!]
  \centering
  \begin{tabular}{lccc}
    \toprule
    \multicolumn{1}{l}{Medium}
    & 
      \multicolumn{1}{l}{%
      $\kappa
      ~
      \left[
      \SI{}{\watt\per(\kelvin~\metre)}
      \right]
      $
      } 
    &
      \multicolumn{1}{l}{%
      $
      \rho
      ~
      \left[
      \SI{}{\kilogram\per\metre^3}
      \right]
      $
      }
    &
      \multicolumn{1}{l}{%
      $
      c_p
      ~
      \left[
      \SI{}{\joule\per(\kilo\gram~\kelvin)}
      \right]
      $
      }      
    \\
    \cmidrule(lr){1-1} 
    \cmidrule(lr){2-2} 
    \cmidrule(lr){3-3}
    \cmidrule(lr){4-4}
    polystyrene & 0.14 & 1000 & 1250
    \\
    water & 0.6 & 1000 & 4185
    \\
    gold &  318 & 19300  & 1860
    \\
    \bottomrule
  \end{tabular}
  \caption{Heat conductivity $\kappa$, density $\rho$ and heat capacity $c_{p}$ used for the numerical calculations in COMSOL \cite{comsol}.}
    \label{tab:simulation_parameters}   
\end{table}


\bibliography{references}
\bibliographystyle{unsrt}

\end{document}